\documentclass[a4paper,fleqn,usenatbib]{mnras}

\usepackage[T1]{fontenc}
\usepackage{ae,aecompl}

\usepackage{cite}
\usepackage{graphicx}	
\usepackage{amsmath}	
\usepackage{amssymb}	
\usepackage{epsfig}
\usepackage{color}
\usepackage{longtable,lscape}

\usepackage{natbib}

\newcommand{\HI}{\mbox{H\,{\sc i}}}

\newcommand{\B}{$B$}
\newcommand{\R}{$R$}
\newcommand{\II}{$I$}

\newcommand{\K}{$K_s$}

\newcommand{\kms}{\mbox{km\,s$^{-1}$}}
\newcommand{\kmst}{km\,s$^{-1}$}
\newcommand{\Jykms}{\mbox{Jy\,km\,s$^{-1}$}}
\newcommand{\Jykmst}{Jy\,km~s$^{-1}$}

\newcommand{\msun}{$M_\odot$}

\newcommand{\lmhi}{$\log(M_{\rmn{HI}})$}

\newcommand{\vlg}{\mbox{${v}_{\rmn{LG}}$}\,}
\newcommand{\vhel}{\mbox{$v_{\rmn{hel}}$}\,}
\newcommand{\vhi}{\mbox{${v}_{\rmn{HI}}$}\,}
\newcommand{\vopt}{\mbox{${v}_{\rmn{opt}}$}\,}

\newcommand{\wfi}{\mbox{$w_{50}$}}
\newcommand{\wtw}{\mbox{$w_{20}$}}

\def\approxlt{\lower.2em\hbox{$\buildrel < \over \sim$}}  
\def\approxgt{\lower.2em\hbox{$\buildrel > \over \sim$}}  

\newcommand{\nan}{Nan\c{c}ay}
\newcommand{\nrt}{NRT}
\newcommand{\mir}{{\tt MIRIAD}}
\newcommand{\sof}{{\tt SoFiA}}
\newcommand{\lar}{{EBHIS}}
\newcommand{\etal}{et~al.}
\newcommand{\cf}{{cf.}\ }
\newcommand{\eg}{{e.g.},\ }         
\newcommand{\ie}{{i.e.},\ }

\definecolor{grey}{rgb}{0.5,0.6,0.7}

\title[EZOA catalogue: H\,I detections in the northern ZoA]
 {EZOA - A catalogue of EBHIS H\,I detected galaxies in the northern Zone of Avoidance }

\author[A.C. Schr\"oder et al.]{
A.C. Schr\"oder,$^{1}$\thanks{E-mail: anja@saao.ac.za}
L. Fl\"oer,$^{2}$
B. Winkel $^{2}$
and J. Kerp $^{2}$
\\
$^{1}$South African Astronomical Observatory, PO Box 9, Observatory 7935, Cape Town, South Africa\\
$^{2}$Argelander Institut f\"ur Astronomie, Auf dem H\"ugel 71, 53121 Bonn, Germany
}

\date{Accepted XXX. Received YYY; in original form ZZZ}

\pubyear{2016}

\begin{document}
\label{firstpage}
\pagerange{\pageref{firstpage}--\pageref{lastpage}}
\maketitle

\begin{abstract}
We present a catalogue of galaxies in the northern Zone of Avoidance (ZoA),
extracted from the shallow version of the blind \HI\ survey with the
Effelsberg 100\,m radio telescope, EBHIS, that has a sensitivity of
23\,mJy\,beam$^{-1}$ at 10.24\,\kms\ velocity resolution. The catalogue
comprises 170 detections in the region $\delta \geq -5\degr$ and $|b| <
6\degr$. About a third of the detections ($N=67$) have not been previously
recorded in \HI . While 29 detections have no discernible counterpart at
any wavelength other than \HI , 48 detections (28\%) have a counterpart
visible on optical or NIR images but are not recorded as such in the
literature. New \HI\ detections were found as close as 7.5\,Mpc (EZOA
J2120+45), and at the edge of the Local Volume, at 10.1\,Mpc, we have found
two previously unknown dwarf galaxies (EZOA J0506+31 and EZOA
J0301+56). Existing large-scale structures crossing the northern ZoA have
been established more firmly by the new detections, with the possibility of
new filaments. We conclude that the high rate of 39\% new \HI\ detections
in the northern ZoA, which has been extensively surveyed with targeted
observations in the past, proves the power of blind \HI\ surveys. The full
EBHIS survey, which will cover the full northern sky with a sensitivity
comparable to the HIPASS survey of the southern sky, is expected to add
many new detections and uncover new structures in the northern ZoA.
\end{abstract}




\section{Introduction}

Surveys of the sky at the 21\,cm wavelength emission of the neutral
hydrogen (\HI ) are a useful means to probe evenly for gas-rich galaxies in
the local universe, including galaxies behind the Galactic plane where dust
and high stellar densities prevent optical surveys from obtaining a full
census of galaxies in a magnitude- or diameter-limited
sample. \HI\ selected galaxy samples are thus free from biases due to
imperfect corrections for Galactic foreground extinctions (\eg
\citealt{riad10}) and stellar densities (\eg \citealt{schroeder19}).

Most \HI\ surveys are pointed observations of a sample of galaxies, usually
selected in the optical or near-infrared (NIR). Systematic blind surveys,
that is, the scanning of the sky to find extragalactic \HI\ emission
independent of prior positional knowledge, became only practicable with the
event of multibeam receivers but are still very time consuming.

The southern sky has been extensively surveyed in \HI\ using the 64\,m
Parkes radio telescope (PKS). The blind survey HIPASS covers all of the
southern sky and part of the northern sky up to $\delta = +25\degr$
(\citealt{meyer04,wong06}) with a sensitivity of $\sim13$\,mJy\,beam$^{-1}$
(at 18\,\kms\ velocity resolution). In addition, the Galactic plane, also
called Zone of Avoidance (ZoA) in extragalactic studies, was surveyed more
deeply with a sensitivity of $\sim6$\,mJy\,beam$^{-1}$ at
27\,\kms\ velocity resolution (HIZOA; \citealt{staveley16,donley05}).

There are no equivalent all-sky surveys of the northern sky. The regions
accessible by the 305\,m Arecibo telescope are covered by ALFALFA
\citep{alfalfa} and the still on-going ALFA\,ZoA
(\citealt{Mcintyre15,henning08}) with a velocity resolution of 9\,\kms\ and
a sensitivity of 2\,mJy\,beam$^{-1}$ and 1\,mJy\,beam$^{-1}$,
respectively. Further north exists the HIJASS survey \citep{lang03}
conducted with the 76\,m Lovell telescope at Jodrell Bank with a published
catalogue covering selected parts of the northern sky at varying
sensitivity ($13-16$\,mJy\,beam$^{-1}$, 18\,\kms\ velocity resolution). The
northern ZoA was partly surveyed by the 25\,m Dwingeloo telescope at a
sensitivity of 40\,mJy\,beam$^{-1}$ \citep{henning98,rivers99}.

The Effelsberg--Bonn \HI\ Survey (EBHIS, \citealt{kerp11},
\citealt{winkel10}), an all-sky survey conducted with the 100\,m Effelsberg
radio telescope, will fill this gap and cover the full northern sky
homogeneously ($\delta \geq -5\degr$) with an ultimate sensitivity of
16\,mJy, comparable to HIPASS. We present here results from the shallow
survey at a sensitivity of 23\,mJy\,beam$^{-1}$ \citep{floer14} at a
velocity resolution of 10\,\kms\ and covering the northern ZoA ($|b| <
6\degr$).

A number of \HI\ surveys are imminent, using the upcoming SKA precursors,
\eg ASKAP \citep{johnston08} in the south and APERTIF \citep{oosterloo10}
and FAST \citep{li13,nan11} in the north; they will cover the sky to much
higher sensitivities, but, as before, only the southern sky will be
homogeneously and continuously covered by ASKAP's all-sky \HI\ survey
WALLABY \citep{koribalski12, duffy12}, whereas the northern sky will only
be surveyed in part, shared between APERTIF and FAST. Hence, EBHIS will
continue to be the only homogeneous northern sky blind \HI\ survey for
quite some time to come.

This paper is organised as follows. Section~\ref{survey} explains the
observations and data reduction; Sec.~\ref{sample} presents the catalogue
and Sec.~\ref{cmatches} the cross-matched data at other wavelengths. The
completeness and reliability of the catalogue is discussed in
Sec.~\ref{compl} and the \HI\ properties of the sample in
Sec.~\ref{hiprop}. How the newly found galaxies fit in with the large scale
structures in the local universe is analysed in Sec.~\ref{lss}. 
Section~\ref{concl} gives a summary. The catalogue, the cross-match table
and the \HI\ profiles are available online.

\section{Observations and data reduction }   \label{survey}

EBHIS \citep{kerp11} is a blind \HI\ survey of the sky north of $\delta =
-5\degr$, conducted with the Effelsberg 100\,m radio telescope. The
Galactic part of the survey ($-600\leq
v_\mathrm{lsr}\leq600~\mathrm{km\,s}^{-1}$) has been published by
\citet{winkel16}.  The here presented catalogue is based on pre-release
\HI\ data\ from the first of two runs, which was finished in April
2013. The second run is on-going and will improve the sensitivity of EBHIS
by about 30\%. The observational parameters of the shallow survey are
summarised in Table~\ref{obstab}.

\begin{table}
\centering
\begin{minipage}{140mm}
\caption{{EBHIS survey parameters} \label{obstab}}
\begin{tabular}{lc}
\hline
Parameter & Value \\
\hline
Coverage & $\delta \geq -5\degr$  \\
Velocity range & $-600$\,\kms$\, < cz < 18,000$\,\kms \\
Velocity resolution & 10.24\,\kms \\
Beam size (FWHP) & $10\farcm8$ \\
Integration time & $\sim\!35$\,s beam$^{-1}$ \\
Cube rms & 23\,mJy beam$^{-1}$  \\
\hline
\end{tabular}
\end{minipage}
\end{table} 

A detailed description of the data reduction techniques applied is given in
\citet{winkel16}; in summary these are: (1) raw data spectra were processed
manually and automatically to search for radio frequency interference
(RFI); (2) a frequency-dependent bandpass (gain) correction was applied;
(3) flux-density or brightness-temperature calibration was based on the
standard IAU source S\,7 \citep{kalberla92}; (4) baselines were subtracted
via a 2D polynomial fitting routine; (5) the side-lobe contribution
(so-called stray radiation) was removed; (6) the resulting, calibrated
spectra were gridded into data cubes.

For the extragalactic part of the data, mostly the same recipe and software
was applied, with two noteworthy differences. First, stray-radiation
removal was not necessary and, second, the baseline subtraction was done
differently.  The 2D polynomial fits to the baseline that were used for the
Galactic part were not practical for the extragalactic data, since a fit to
the full velocity range of $-600\leq v_\mathrm{lsr}\leq18,000$\kms\ would
require a much higher order polynomial. A possible solution was to fit
chunks of data at a time, that is, about 1000 spectral channels interleaved
by 500 channels. This was found to be relatively time consuming, but
yielded results that are comparable to the Galactic EBHIS data
quality. However, with a median spectrum computed over a larger field (\eg
$10\degr\times10\degr$), small-scale residual ripples may appear in the
baseline.

It was therefore decided to follow a slightly different approach for the
extragalactic baseline removal. Before any polynomial fitting was
performed, a median spectrum of the raw calibrated spectra was calculated
for each subscan, feed and polarisation channel which was then subtracted
from each individual spectrum. This effectively removed the small-scale
ripples.  For the Galactic velocity interval such an approach would not
have been possible because the \HI\ line is present in every single
spectrum. For the extragalactic part, however, only a very small fraction
of the data contains \HI\ line signals and sufficient ``baseline",
surrounding the sources of interest, is available. Since continuum sources
usually have an impact on the baseline shape \citep[see,
  e.g.,][]{winkel16}, spectra such affected were neglected for the
median-calculation. As also explained by \citet{winkel16}, the baselines
have a noticeable time-dependency. Therefore, an additional low-order 2D
polynomial fit was performed and subtracted to account for the temporal
evolution.

Finally, to ease working with the spectra, data cubes of size
$12\degr\times12\degr$ (with an overlap of $\Delta\ell = 1\degr$ on each
side) are produced and spectrally binned with a factor of eight, leading to
an effective spectral resolution of 10.24\,\kms , to reduce the data
volume.

\section{Sample Compilation and HI Parameterisation }  \label{sample}

For the search in the ZoA, cubes with a velocity range of $-500\leq cz
\leq12,500$\,\kms\ and $12\degr\times12\degr$ were produced with a spatial
overlap of $1\degr$ in Galactic longitude. We adopted an upper velocity
limit since we do not expect detections much above $10,000$\,\kms\ at this
sensitivity. The velocity axis in the cubes were converted from LSRK
(kinematic local standard of rest) to \vhel \citep{kerr86}, and the
intensity scale from Kelvin to Jansky. To ease the problem of strong
baseline variations caused by continuum sources, we worked with two sets of
cubes: the `normal' ones, as are, and a set where a spline-based baseline
was subtracted along each line of sight to remove the strongest baseline
variations. This, however, caused troughs around the brighter \HI\ sources
which may affect their visibility to the eye.

Each ZoA cube in both sets was visually inspected using the visualisation
package {\tt KARMA} \citep{gooch96}.  A comparison of the ensuing source
lists from the two sets showed that, as expected, bright detections
affected the baseline-splinefit and were often harder to see in the
baseline-subtracted cubes, but the latter made it easier to find galaxies
where strong baseline ripples occur.  The resulting, comprehensive source
list was re-checked and quality flags applied (based on signal-to-noise
ratio, presence of baseline wiggles and visual appearance of the
detection). Only detections deemed reliable were retained for the catalogue
where those with a signal-to-noise ratio below 6 are considered
marginal. Reliability determinations were based on independent judgements
for two slices (XZ and ZY\footnote{In the \HI\ data cubes, X and Y are the
two coordinates and Z is the velocity axis.}) of each set of cubes (using
the appearance in the image and the individual line-of-sight profiles) as
well as the final fitted profile; where in doubt we compared detections of
similar quality to ensure homogeneity in the judgements.  Detections which
did not make this cut (\eg with a too low signal-to-noise ratio or not
convincingly real) were retained for comparison with the upcoming full
EBHIS survey.

\HI\ parameters were determined using the programme {\tt MBSPECT} within
the \mir\ package \citep{sault95}. Zeroth moment maps were made for each
detection using the non-baseline subtracted cubes, and positions were
determined using a Gaussian fit. A one-dimensional Hanning-smoothed
spectral profile was obtained by calculating the weighted sum of the
emission at the resulting position. The spectrum was visually inspected and
a low-order polynomial was fitted to the emission-free channels and
subtracted. An integration over the channels containing the \HI\ emission
of the baseline-subtracted spectrum resulted in the total flux. The
heliocentric velocity was taken to be the mean of the velocity values at
the 50\% mark of the peak flux density of the profile. Line widths at both
the 50\% as well as the 20\% level of the peak flux density were determined
using a width-maximising algorithm.

Errors were calculated using the formalism presented in
\citet{koribalski04}.  The errors on velocities and line widths depend,
among others, on the steepness of the profile edges and thus on the line
width at 20\% of peak flux density; in case this value could not be
determined (where the signal was too close to the noise level, $N=38$
profiles), we used the median steepness derived from our sample,
($w_{20}-w_{50}) = 24.5\,$\kms , to calculate the errors for $w_{50}$ and
$v_{\rm hel}$.

\begin{figure*}
\centering
\includegraphics[width=0.95\textwidth]{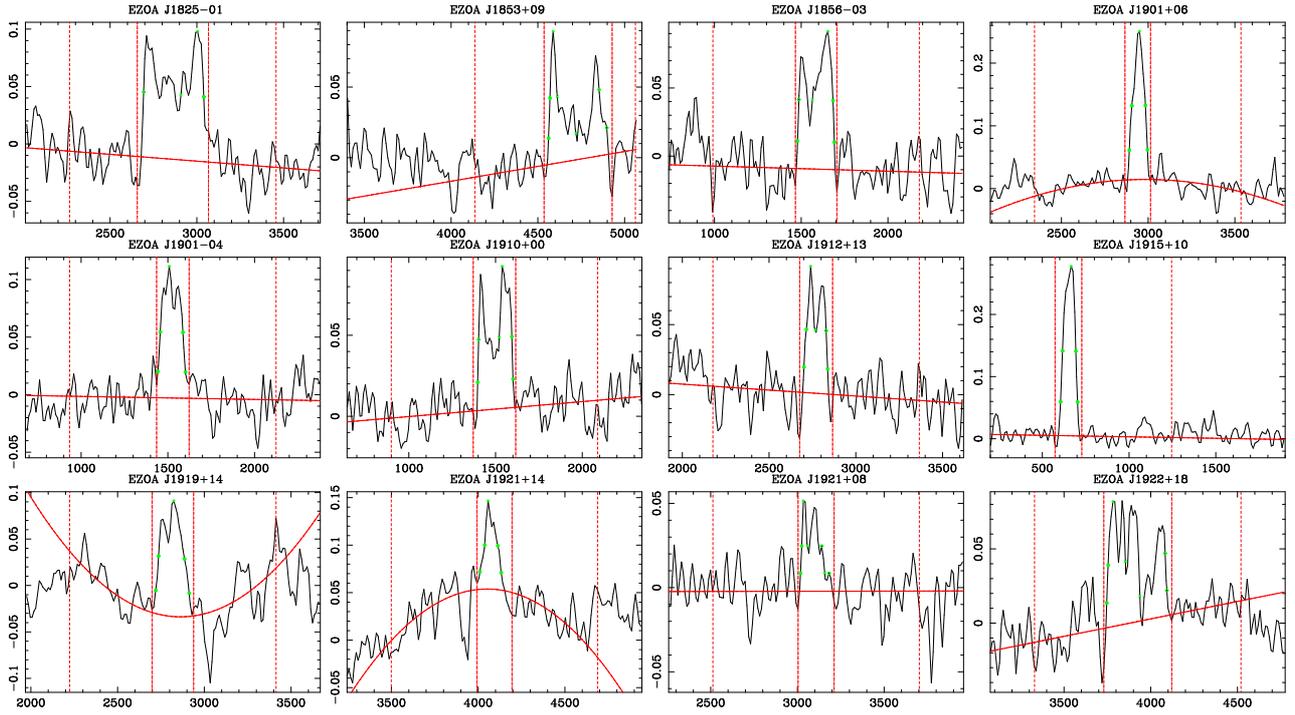} 
\caption{Example \HI\ spectra of the newly detected galaxies in the EZOA
  blind survey (Flux densities, in Jy\,\kms , versus radial heliocentric
  velocity, in \kms ); all the spectra are available online. Low order
  baselines (indicated by the solid line) are fitted, excluding the
  detections themselves (which are bracketed by the dash-dot vertical
  lines) and excluding the low and high-velocity edges to the left and
  right of the dashed vertical lines, respectively. 20\% and 50\% profile
  markers are visible.
}
\label{specplot}
\end{figure*}

We have detected 170 \HI\ sources. The catalogue and the \HI\ profile plots
are available online; Table~\ref{tabdets} and Fig.~\ref{specplot} give the
first 12 detections and profiles, respectively, as an example. On each
profile, the peak as well as the 50\% and 20\% levels are noted with
dots. Vertical lines indicate the spectral ranges used for baseline
subtraction, and the linear or polynomial fit is shown. Table~\ref{tabdets}
lists \HI\ parameters and derived quantities for the galaxies in the
following columns:

Columns (1): Source name.

Columns (2a and 2b): Equatorial coordinates (J2000.0) of the fitted
position.

Columns (3a and 3b): Galactic coordinates.

Column (4): Reddening $E(B-V)$ as derived from the IRAS/DIRBE maps
\citep{schlegel98} and corrected with a factor of 0.86 as derived by
\citet{schlafly11}.

Column (5): Heliocentric velocity and error, in \kms .

Column (6): Velocity width at 50\% of peak flux density and associated
error, in \kms .

Column (7): Velocity width at 20\% of peak flux density and associated
error, in \kms .


Column (8): \HI\ flux integral and associated error, in Jy\,\kms .

Column (9): Signal-to-noise ration SNR using the peak flux.

Column (10): Velocity of the galaxy, in \kms , corrected to the Local Group
frame of reference via:
$$ v_{\rm LG} = v_{\rm hel} + 300 \sin \ell \cos b $$

Column (11): Distance to the galaxy in Megaparsec, based on $v_{\rm LG}$
and H$_0$ = 75\,\kms Mpc$^{-1}$.

Column (12): Logarithm of the \HI\ mass.

\section{Multiwavelength counterparts  }   \label{cmatches}

The search for counterparts (\ie at wavelengths other than \HI ) was done
using the following online literature databases and optical and NIR images:
The NASA/IPAC Extragalactic Database
(NED)\footnote{http://ned.ipac.caltech.edu},
SIMBAD\footnote{http://simbad.u-strasbg.fr/simbad/}, the SuperCOSMOS Sky
Surveys\footnote{http://www-wfau.roe.ac.uk/sss/} (\B -band), the Digitized
Sky Surveys
(DSS)\footnote{http://www3.cadc-ccda.hia-iha.nrc-cnrc.gc.ca/en/dss/} (\R -
and \II -band),
2MASS\footnote{http://irsa.ipac.caltech.edu/applications/2MASS/} (\K
-band), UKIDSS\footnote{http://surveys.roe.ac.uk/wsa/} and
VISTA\footnote{http://horus.roe.ac.uk/vsa/} (both preferably \K -band).
The search procedure is described in detail in \citet{staveley16} with the
main difference that the search radius (which depends on the spatial
resolution of the \HI\ data) was set to $3\farcm0$.

The most likely counterpart was identified based on the \HI\ parameters,
the appearance of the galaxy on the images and the extinction information:
Line widths were compared with the inclination and morphological type of the
galaxies; the \HI\ flux was compared with apparent brightness, size and
optical velocity measurements (where available), taking into account the
obscuring effect of the local extinction.  Table~\ref{tabcross} lists 16
detections and their counterparts (where available) as an example, the rest
is available online only.  There are three sub-sections:
Table~\ref{tabcross}a gives the EZOA detections with either a single or no
counterpart. Table~\ref{tabcross}b lists detections where more than one
galaxy is assumed to contribute to the \HI\ profile (note that in the case
of EZOA J0440$+$49A and EZOA J0440$+$49B the profiles are confused but the
detections could be fitted separately -- they are thus listed in
Table~\ref{tabcross}a). Table~\ref{tabcross}c presents those cases where
more than one candidate was found but, judged by the profile, only {\it one
  of them} is the likely counterpart. The columns are as follows:

Column (1): Source name as in Table~\ref{tabdets}.

Columns (2a and 2b): Galactic coordinates of the \HI\ detection.

Column (3): Distance to the \HI\ galaxy in Megaparsec, as in
Table~\ref{tabdets}.

Column (4): Logarithm of the \HI\ mass, as in Table~\ref{tabdets}.

Column (5): Extinction in the \B -band, converted from $E(B-V)$ given in
Table~\ref{tabdets} using $R_{\rm B}=4.14$. A star denotes an extinction
value deemed to be uncertain during the search (e.g., due to high spatial
variability), whereas a question mark indicates the possibility that the
extinction value might be erroneous.

Column (6): Classification of the counterpart; `d' = definite, `p' =
probable, `a' = ambiguous, `c' = confused candidate; `--' = no candidate.

Column (7): Flags for counterparts in major catalogues: `I' stands for IRAS
Point Source Catalog \citep{iras}, `M' for 2MASX \citep{jarrett00b}, `H'
for the HIPASS catalogues (South, \citealt{meyer04} and North,
\citealt{wong06}), `J' for the HI Jodrell All Sky Survey (HIJASS;
\citealt{lang03}), `S' for the \HI\ Parkes ZOA Shallow Survey (HIZSS,
\citealt{henning00a}), and `Z' for HIZOA publications
(\citealt{juraszek00,donley05,staveley16}).

Column (8): Type of velocity measurement in the literature (from NED or
HyperLEDA\footnote{http://leda.univ-lyon1.fr/}): `o' = optical, `h' = \HI .

Column (9): Source `l' for coordinates from the literature: `N' is listed
in NED, `S' is listed in SIMBAD. The flag `c' stands for coordinates
measured on DSS or NIR images (note that some published coordinates were
not centred properly so we give the measured ones).

Column (10): Note in the appendix.

Columns (11a and 11b): Equatorial coordinates (J2000.0) of the counterpart.

Column (12): Distance between the \HI\ fitted position and the counterpart
position in arcminutes.

Column (13): One name in the literature of an optical- or NIR-known
counterpart, in this order of preference: NGC, IC, UGC, ESO, CGMW, WEIN,
2MASS, others.

\onecolumn
%
{\scriptsize
\begin{longtable}{
l@{\extracolsep{0.mm}}                                      r@{\extracolsep{2.mm}}                                   r@{\extracolsep{4.mm}}                         
 r@{\extracolsep{1.mm}}                                     r@{\extracolsep{-1.mm}}                                  r@{\extracolsep{2.mm}}                        
 r@{\extracolsep{1.mm}}                                     r@{\extracolsep{1.mm}}                                   r@{\extracolsep{3.mm}}                       
 r@{\extracolsep{1.mm}}                                    r@{\extracolsep{1.mm}}                                   r@{\extracolsep{-1.mm}}                                  
 r@{\extracolsep{1.mm}}                                     r}
\caption{EZOA \HI\ detections with measured and derived parameters \label{tabdets}} \\
\hline\hline 
\noalign{\smallskip}
EZOA ID                                                       & \multicolumn{1}{c@{\extracolsep{2.mm}}}{RA}            & \multicolumn{1}{c@{\extracolsep{4.mm}}}{Dec}      
 & \multicolumn{1}{c@{\extracolsep{1.mm}}}{$l$}               & \multicolumn{1}{c@{\extracolsep{-1.mm}}}{$b$}          & \multicolumn{1}{r@{\extracolsep{2.mm}}}{$E(B-V)$}   
 & \multicolumn{1}{c@{\extracolsep{1.mm}}}{$v_{\rm hel}$}     & \multicolumn{1}{c@{\extracolsep{1.mm}}}{$w_{50}$}      & \multicolumn{1}{c@{\extracolsep{3.mm}}}{$w_{20}$} 
 & \multicolumn{1}{c@{\extracolsep{-1.mm}}}{Flux}             & \multicolumn{1}{c@{\extracolsep{-1.mm}}}{SNR}          & \multicolumn{1}{c@{\extracolsep{-1.mm}}}{$v_{\rm LG}$} 
 & \multicolumn{1}{c@{\extracolsep{1.mm}}}{$D$}               & \multicolumn{1}{c@{\extracolsep{1.mm}}}{$\log M_{\rm HI}$} \\
                                                              & \multicolumn{2}{c@{\extracolsep{2.mm}}}{J2000.0}        
 & \multicolumn{1}{c@{\extracolsep{1.mm}}}{[deg]}             & \multicolumn{1}{c@{\extracolsep{-1.mm}}}{[deg]}        &                                                   
 & \multicolumn{1}{c@{\extracolsep{1.mm}}}{[\kmst]}           & \multicolumn{1}{c@{\extracolsep{1.mm}}}{[\kmst]}       & \multicolumn{1}{c@{\extracolsep{3.mm}}}{[\kmst]}  
 & \multicolumn{1}{r@{\extracolsep{-1.mm}}}{[\Jykmst]}        &                                                        & \multicolumn{1}{c@{\extracolsep{-1.mm}}}{[\kmst]}      
 & \multicolumn{1}{c@{\extracolsep{1.mm}}}{[Mpc]}             & \multicolumn{1}{c@{\extracolsep{1.mm}}}{[\msun]} \\
\multicolumn{1}{c}{(1)}                                       & \multicolumn{1}{c@{\extracolsep{2.mm}}}{(2a)}          & \multicolumn{1}{c@{\extracolsep{4.mm}}}{(2b)}      
 & \multicolumn{1}{c@{\extracolsep{1.mm}}}{(3a)}              & \multicolumn{1}{c@{\extracolsep{-1.mm}}}{(3b)}         & \multicolumn{1}{c@{\extracolsep{2.mm}}}{(4)}      
 & \multicolumn{1}{c@{\extracolsep{1.mm}}}{(5)}               & \multicolumn{1}{c@{\extracolsep{1.mm}}}{(6)}           & \multicolumn{1}{c@{\extracolsep{3.mm}}}{(7)}      
 & \multicolumn{1}{c@{\extracolsep{-1.mm}}}{(8)}              & \multicolumn{1}{c@{\extracolsep{-1.mm}}}{(9)}          & \multicolumn{1}{c@{\extracolsep{1.mm}}}{(10)}    
 & \multicolumn{1}{c@{\extracolsep{1.mm}}}{(11)}              & \multicolumn{1}{c@{\extracolsep{1.mm}}}{(12)}    \\
\noalign{\smallskip}
\hline
\noalign{\smallskip}
\endfirsthead
\caption{--Continued.}\\
\hline\hline
\noalign{\smallskip}
EZOA ID                                                       & \multicolumn{1}{c@{\extracolsep{2.mm}}}{RA}            & \multicolumn{1}{c@{\extracolsep{4.mm}}}{Dec}      
 & \multicolumn{1}{c@{\extracolsep{1.mm}}}{$l$}               & \multicolumn{1}{c@{\extracolsep{-1.mm}}}{$b$}          & \multicolumn{1}{r@{\extracolsep{2.mm}}}{$E(B-V)$}   
 & \multicolumn{1}{c@{\extracolsep{1.mm}}}{$v_{\rm hel}$}     & \multicolumn{1}{c@{\extracolsep{1.mm}}}{$w_{50}$}      & \multicolumn{1}{c@{\extracolsep{3.mm}}}{$w_{20}$} 
 & \multicolumn{1}{c@{\extracolsep{-1.mm}}}{Flux}             & \multicolumn{1}{c@{\extracolsep{-1.mm}}}{SNR}          & \multicolumn{1}{c@{\extracolsep{-1.mm}}}{$v_{\rm LG}$} 
 & \multicolumn{1}{c@{\extracolsep{1.mm}}}{$D$}               & \multicolumn{1}{c@{\extracolsep{1.mm}}}{$\log M_{\rm HI}$} \\
                                                              & \multicolumn{2}{c@{\extracolsep{2.mm}}}{J2000.0}        
 & \multicolumn{1}{c@{\extracolsep{1.mm}}}{[deg]}             & \multicolumn{1}{c@{\extracolsep{-1.mm}}}{[deg]}        &                                                   
 & \multicolumn{1}{c@{\extracolsep{1.mm}}}{[\kmst]}           & \multicolumn{1}{c@{\extracolsep{1.mm}}}{[\kmst]}       & \multicolumn{1}{c@{\extracolsep{3.mm}}}{[\kmst]}  
 & \multicolumn{1}{r@{\extracolsep{-1.mm}}}{[\Jykmst]}        &                                                        & \multicolumn{1}{c@{\extracolsep{-1.mm}}}{[\kmst]}      
 & \multicolumn{1}{c@{\extracolsep{1.mm}}}{[Mpc]}             & \multicolumn{1}{c@{\extracolsep{1.mm}}}{[\msun]} \\
\multicolumn{1}{c}{(1)}                                       & \multicolumn{1}{c@{\extracolsep{2.mm}}}{(2a)}          & \multicolumn{1}{c@{\extracolsep{4.mm}}}{(2b)}      
 & \multicolumn{1}{c@{\extracolsep{1.mm}}}{(3a)}              & \multicolumn{1}{c@{\extracolsep{-1.mm}}}{(3b)}         & \multicolumn{1}{c@{\extracolsep{2.mm}}}{(4)}      
 & \multicolumn{1}{c@{\extracolsep{1.mm}}}{(5)}               & \multicolumn{1}{c@{\extracolsep{1.mm}}}{(6)}           & \multicolumn{1}{c@{\extracolsep{3.mm}}}{(7)}      
 & \multicolumn{1}{c@{\extracolsep{-1.mm}}}{(8)}              & \multicolumn{1}{c@{\extracolsep{-1.mm}}}{(9)}          & \multicolumn{1}{c@{\extracolsep{1.mm}}}{(10)}    
 & \multicolumn{1}{c@{\extracolsep{1.mm}}}{(11)}              & \multicolumn{1}{c@{\extracolsep{1.mm}}}{(12)}    \\
\noalign{\smallskip}
\hline
\noalign{\smallskip}
\endhead
\noalign{\smallskip}
\hline
\endfoot
J1825$-$01   & 18 25 01.7 &  $-01$ 30 13 &   28.57 & $   5.18$ &  1.38 &   2867$\pm$\phantom{0}7 &  347$\pm$\phantom{}13 &  \multicolumn{1}{c}{\phantom{00}\ldots}&  27.2$\pm$\phantom{0}4.3 &   7.3 &   3010 &  40.1 & 10.01 \\ 
J1853$+$09   & 18 53 52.2 &  $+09$ 49 41 &   41.97 & $   3.93$ &  0.79 &   4711$\pm$\phantom{0}7 &  283$\pm$\phantom{}13 &                   336$\pm$\phantom{}20 &  12.4$\pm$\phantom{0}1.8 &  10.7 &   4911 &  65.5 & 10.10 \\ 
J1856$-$03   & 18 56 03.1 &  $-03$ 14 28 &   30.57 & $  -2.50$ &  1.20 &   1585$\pm$\phantom{0}5 &  196$\pm$\phantom{}10 &                   212$\pm$\phantom{}15 &  14.5$\pm$\phantom{0}2.9 &   7.6 &   1737 &  23.2 &  9.26 \\ 
J1901$+$06   & 19 01 34.4 &  $+06$ 51 48 &   40.19 & $   0.88$ &  4.72 &   2945$\pm$\phantom{0}2 &   78$\pm$\phantom{0}4 &                   105$\pm$\phantom{0}6 &  17.0$\pm$\phantom{0}1.5 &  24.2 &   3139 &  41.8 &  9.85 \\ 
J1901$-$04   & 19 01 46.2 &  $-04$ 29 27 &   30.10 & $  -4.34$ &  0.68 &   1521$\pm$\phantom{0}5 &  131$\pm$\phantom{}11 &                   159$\pm$\phantom{}16 &  13.2$\pm$\phantom{0}2.3 &   9.6 &   1671 &  22.3 &  9.19 \\ 
J1910$+$00   & 19 10 22.7 &  $+00$ 31 05 &   35.56 & $  -3.98$ &  0.59 &   1498$\pm$\phantom{0}4 &  191$\pm$\phantom{0}8 &                   204$\pm$\phantom{}12 &  11.3$\pm$\phantom{0}2.2 &   8.4 &   1672 &  22.3 &  9.12 \\ 
J1912$+$13   & 19 12 42.2 &  $+13$ 24 28 &   47.26 & $   1.46$ &  2.51 &   2772$\pm$\phantom{0}5 &  114$\pm$\phantom{}10 &                   136$\pm$\phantom{}15 &   8.2$\pm$\phantom{0}1.7 &   9.1 &   2992 &  39.9 &  9.49 \\ 
J1915$+$10   & 19 15 00.5 &  $+10$ 16 53 &   44.76 & $  -0.49$ & 10.56 &    655$\pm$\phantom{0}1 &   77$\pm$\phantom{0}3 &                    98$\pm$\phantom{0}4 &  20.3$\pm$\phantom{0}1.3 &  31.7 &    866 &  11.5 &  8.80 \\ 
J1919$+$14   & 19 19 53.2 &  $+14$ 09 02 &   48.73 & $   0.27$ &  7.09 &   2811$\pm$\phantom{0}9 &  150$\pm$\phantom{}17 &                   193$\pm$\phantom{}26 &  16.8$\pm$\phantom{0}3.6 &   7.3 &   3036 &  40.5 &  9.81 \\ 
J1921$+$14   & 19 21 38.8 &  $+14$ 53 08 &   49.58 & $   0.23$ &  4.70 &   4075$\pm$\phantom{}11 &   75$\pm$\phantom{}21 &                   123$\pm$\phantom{}32 &   7.1$\pm$\phantom{0}2.4 &   6.2 &   4303 &  57.4 &  9.74 \\ 
J1921$+$08   & 19 21 54.2 &  $+08$ 18 55 &   43.81 & $  -2.91$ &  0.90 &   3082$\pm$\phantom{}11 &  115$\pm$\phantom{}22 &                   164$\pm$\phantom{}33 &   5.2$\pm$\phantom{0}1.5 &   6.2 &   3289 &  43.9 &  9.37 \\ 
J1922$+$18   & 19 22 47.3 &  $+18$ 45 13 &   53.12 & $   1.82$ &  2.81 &   3919$\pm$\phantom{0}4 &  329$\pm$\phantom{0}8 &                   347$\pm$\phantom{}12 &  17.2$\pm$\phantom{0}2.1 &  10.2 &   4159 &  55.5 & 10.10 \\ 
\multicolumn{12}{l}{$\cdots$} \\   
\end{longtable}
}
{\scriptsize
\begin{longtable}{
l@{\extracolsep{-1.mm}}                            r@{\extracolsep{1.mm}}                                       r@{\extracolsep{0.mm}}
 r@{\extracolsep{-1.mm}}                           r@{\extracolsep{1.mm}}                                       r@{\extracolsep{1.mm}}
 c@{\extracolsep{2.mm}}                            c@{\extracolsep{0.mm}}                                       c@{\extracolsep{0.mm}}
                                                   c@{\extracolsep{0.mm}}                                       c@{\extracolsep{0.mm}}
 c@{\extracolsep{2.mm}}                            c@{\extracolsep{-2.mm}}                                      c@{\extracolsep{1.mm}}
 c@{\extracolsep{-2.mm}}                           c@{\extracolsep{2.mm}}                                       c@{\extracolsep{2.mm}}
 l@{\extracolsep{2.mm}}                            l@{\extracolsep{1.mm}}                                       r@{\extracolsep{4.mm}}
 l}
\caption{Crossmatches of the EZOA HI detections \label{tabcross}} \\
\hline\hline 
\noalign{\smallskip}
\multicolumn{1}{c@{\extracolsep{-1.mm}}}{EZOA ID}  & \multicolumn{1}{c@{\extracolsep{1.mm}}}{$l$}               & \multicolumn{1}{c@{\extracolsep{0.mm}}}{$b$}   
 & \multicolumn{1}{c@{\extracolsep{-1.mm}}}{$D$}   & \multicolumn{1}{c@{\extracolsep{1.mm}}}{$\log M_{\rm HI}$} & \multicolumn{1}{c@{\extracolsep{1.mm}}}{$A_B$}             
 & class                                           & I                                                          & M
                                                   & H                                                          & S
 & Z                                               & o                                                          & h 
 & l                                               & c                                                          & Note  
 & \multicolumn{1}{c@{\extracolsep{2.mm}}}{RA}     & \multicolumn{1}{c@{\extracolsep{1.mm}}}{Dec}               & \multicolumn{1}{c@{\extracolsep{4.mm}}}{$d_{\rm sep}$}   
 & \multicolumn{1}{l@{\extracolsep{1.mm}}}{Name}   \\  
                                                  & \multicolumn{1}{c@{\extracolsep{1.mm}}}{[deg]}             & \multicolumn{1}{c@{\extracolsep{0.mm}}}{[deg]} 
 & \multicolumn{1}{c@{\extracolsep{-1.mm}}}{[Mpc]} & \multicolumn{1}{c@{\extracolsep{1.mm}}}{[\msun]}           & \multicolumn{1}{c@{\extracolsep{1.mm}}}{[mag]}           
 &                                                 &                                                            &  
                                                   &                                                            &  
 &                                                 &                                                            &
 &                                                 &                                                            & 
 & \multicolumn{2}{c@{\extracolsep{1.mm}}}{J2000.0}                                                             & \multicolumn{1}{c@{\extracolsep{4.mm}}}{[$\prime$]} 
 &                                                 \\
\multicolumn{1}{c}{(1)} & \multicolumn{1}{c@{\extracolsep{1.mm}}}{(2a)}              & \multicolumn{1}{c@{\extracolsep{0.mm}}}{(2b)}   
 & \multicolumn{1}{c@{\extracolsep{-1.mm}}}{(3)}   & \multicolumn{1}{c@{\extracolsep{1.mm}}}{(4)}               & \multicolumn{1}{c@{\extracolsep{1.mm}}}{(5)}  
 & \multicolumn{1}{c@{\extracolsep{-1.mm}}}{(6)}   & \multicolumn{5}{c@{\extracolsep{0.mm}}}{(7)}                
                                                   & \multicolumn{1}{c@{\extracolsep{0.mm}}}{(8)}               & 
 & \multicolumn{1}{c@{\extracolsep{-1.mm}}}{(9)}   &                                                            & (10) 
 & \multicolumn{1}{c@{\extracolsep{2.mm}}}{(11a)}  & \multicolumn{1}{c@{\extracolsep{1.mm}}}{(11b)}             & \multicolumn{1}{c@{\extracolsep{4.mm}}}{(12)}  
 & \multicolumn{1}{l@{\extracolsep{1.mm}}}{(13)} \\
\noalign{\smallskip}
\hline
\noalign{\smallskip}
\endfirsthead
\caption{--Continued.}\\
\hline\hline
\noalign{\smallskip}
\multicolumn{1}{c@{\extracolsep{-1.mm}}}{EZOA ID}  & \multicolumn{1}{c@{\extracolsep{1.mm}}}{$l$}               & \multicolumn{1}{c@{\extracolsep{0.mm}}}{$b$}   
 & \multicolumn{1}{c@{\extracolsep{-1.mm}}}{$D$}   & \multicolumn{1}{c@{\extracolsep{1.mm}}}{$\log M_{\rm HI}$} & \multicolumn{1}{c@{\extracolsep{1.mm}}}{$A_B$}             
 & class                                           & I                                                          & M
                                                   & H                                                          & S
 & Z                                               & o                                                          & h 
 & l                                               & c                                                          & Note  
 & \multicolumn{1}{c@{\extracolsep{2.mm}}}{RA}     & \multicolumn{1}{c@{\extracolsep{1.mm}}}{Dec}               & \multicolumn{1}{c@{\extracolsep{4.mm}}}{$d_{\rm sep}$}   
 & \multicolumn{1}{l@{\extracolsep{1.mm}}}{Name}   \\  
                                                  & \multicolumn{1}{c@{\extracolsep{1.mm}}}{[deg]}             & \multicolumn{1}{c@{\extracolsep{0.mm}}}{[deg]} 
 & \multicolumn{1}{c@{\extracolsep{-1.mm}}}{[Mpc]} & \multicolumn{1}{c@{\extracolsep{1.mm}}}{[\msun]}           & \multicolumn{1}{c@{\extracolsep{1.mm}}}{[mag]}           
 &                                                 &                                                            &  
                                                   &                                                            &  
 &                                                 &   
 &                                                 &                                                            & 
 & \multicolumn{2}{c@{\extracolsep{1.mm}}}{J2000.0}                                                             & \multicolumn{1}{c@{\extracolsep{4.mm}}}{[$\prime$]} 
 &                                                 \\
\multicolumn{1}{c}{(1)} & \multicolumn{1}{c@{\extracolsep{1.mm}}}{(2a)}              & \multicolumn{1}{c@{\extracolsep{0.mm}}}{(2b)}   
 & \multicolumn{1}{c@{\extracolsep{-1.mm}}}{(3)}   & \multicolumn{1}{c@{\extracolsep{1.mm}}}{(4)}               & \multicolumn{1}{c@{\extracolsep{1.mm}}}{(5)}  
 & \multicolumn{1}{c@{\extracolsep{-1.mm}}}{(6)}   & \multicolumn{5}{c@{\extracolsep{0.mm}}}{(7)}                
                                                   & \multicolumn{1}{c@{\extracolsep{0.mm}}}{(8)}               & 
 & \multicolumn{1}{c@{\extracolsep{-1.mm}}}{(9)}   &                                                            & (10) 
 & \multicolumn{1}{c@{\extracolsep{2.mm}}}{(11a)}  & \multicolumn{1}{c@{\extracolsep{1.mm}}}{(11b)}             & \multicolumn{1}{c@{\extracolsep{4.mm}}}{(12)}  
 & \multicolumn{1}{l@{\extracolsep{1.mm}}}{(13)} \\
\noalign{\smallskip}
\hline
\noalign{\smallskip}
\endhead
\noalign{\smallskip}
\hline
\endfoot
\multicolumn{13}{l}{} \\ 
\multicolumn{13}{l}{(a) HI detections with single cross-matches:} \\
\multicolumn{13}{l}{} \\   
J1825$-$01  &  28.57 & $  5.18$ &  40.1  & 10.01 &  5.7\phantom{*} & d  & --&--& H&--& Z&  --& h&  --&c  & -- & 18 25 00.1                            &                         $-$01 28 33   &  1.6   &  \ldots                   \\
J1853$+$09  &  41.97 & $  3.93$ &  65.5  & 10.10 &  3.3\phantom{*} & d  & --& M& H&--& Z&  --& h&   N&c  & -- & 18 53 47.5                            &               \phantom{$-$}09 51 14   &  2.0   &  2MASX J18534771+0951131  \\
J1856$-$03  &  30.57 & $ -2.50$ &  23.2  &  9.26 &  5.0\phantom{*} & -- & --&--& H& S& Z&  --& h&  --&-- & n  & \multicolumn{1}{c}{\phantom{0}\ldots} & \multicolumn{1}{c}{\phantom{0}\ldots} & \ldots &  \ldots                   \\
J1901$+$06  &  40.19 & $  0.88$ &  41.8  &  9.85 & 19.5*           & d  & --&--& H&--& Z&  --& h&  --&c  & -- & 19 01 35.2                            &               \phantom{$-$}06 51 30   &  0.4   &  \ldots                   \\
J1901$-$04  &  30.10 & $ -4.34$ &  22.3  &  9.19 &  2.8*           & -- & --&--& H& S& Z&  --& h&  --&-- & -- & \multicolumn{1}{c}{\phantom{0}\ldots} & \multicolumn{1}{c}{\phantom{0}\ldots} & \ldots &  \ldots                   \\
J1910$+$00  &  35.56 & $ -3.98$ &  22.3  &  9.12 &  2.4\phantom{*} & d  & --&--& H& S& Z&  --& h&  --&c  & -- & 19 10 24.9                            &               \phantom{$-$}00 32 21   &  1.4   &  \ldots                   \\
J1912$+$13  &  47.26 & $  1.46$ &  39.9  &  9.49 & 10.4\phantom{*} & -- & --&--& H&--& Z&  --& h&  --&-- & -- & \multicolumn{1}{c}{\phantom{0}\ldots} & \multicolumn{1}{c}{\phantom{0}\ldots} & \ldots &  \ldots                   \\
J1915$+$10  &  44.76 & $ -0.49$ &  11.5  &  8.80 & 43.7\phantom{*} & -- & --&--& H&--& Z&  --& h&  --&-- & -- & \multicolumn{1}{c}{\phantom{0}\ldots} & \multicolumn{1}{c}{\phantom{0}\ldots} & \ldots &  \ldots                   \\
J1919$+$14  &  48.73 & $  0.27$ &  40.5  &  9.81 & 29.4\phantom{*} & -- & --&--& H&--& Z&  --& h&  --&-- & n  & \multicolumn{1}{c}{\phantom{0}\ldots} & \multicolumn{1}{c}{\phantom{0}\ldots} & \ldots &  \ldots                   \\
J1921$+$14  &  49.58 & $  0.23$ &  57.4  &  9.74 & 19.5*           & p  & --&--& H&--& Z&  --& h&  --&c  & n  & 19 21 35.1                            &               \phantom{$-$}14 50 19   &  2.9   &  \ldots                   \\
J1921$+$08  &  43.81 & $ -2.91$ &  43.9  &  9.37 &  3.7\phantom{*} & d  & --&--& H&--& Z&  --& h&  --&c  & -- & 19 22 00.2                            &               \phantom{$-$}08 18 24   &  1.6   &  \ldots                   \\
J1922$+$18  &  53.12 & $  1.82$ &  55.5  & 10.10 & 11.6\phantom{*} & d  & --&--& H&--&--&  --& h&  --&c  & -- & 19 22 45.8                            &               \phantom{$-$}18 42 40   &  2.6   &  \ldots                   \\
\multicolumn{12}{l}{$\cdots$} \\   
\noalign{\smallskip}                                                                                                                                 
\multicolumn{12}{l}{(b) HI-detections with more than one cross-match: } \\
\noalign{\smallskip}                                                                                                                                 
J1929$+$08  &  44.43 & $ -4.67$ &  44.2 &  9.85 &  1.6\phantom{*} & d  &  I&--& H&--& Z&  --& h&   S&-- & n  & 19 29 19.8                            &               \phantom{$-$}08 02 42   &  1.0   &  DSH J1929.3+0802         \\
     $ $    &        & $      $ &       &       &     \phantom{*} & c  & --&--& H&--& Z&  --& h&   S&c  & n  & 19 29 19.9                            &               \phantom{$-$}08 04 46   &  3.0   &  DSH J1929.3+0804         \\
J2143$+$46  &  92.71 & $ -4.97$ &  46.8 & 10.03 &  1.4\phantom{*} & d  &  I& M&--&--&--&   o& h&   N&-- & n  & 21 43 54.08                           &               \phantom{$-$}46 37 04.7 &  0.5   &  UGC 11802                \\
     $ $    &        & $      $ &       &       &     \phantom{*} & c  &  I& M&--&--&--&   o& h&   N&-- & n  & 21 44 13.47                           &               \phantom{$-$}46 37 16.9 &  3.1   &  UGC 11806                \\
\multicolumn{12}{l}{$\cdots$} \\   
\noalign{\smallskip}                                                                                                                                 
\multicolumn{12}{l}{(c) HI detections with ambiguous cross-matches:} \\
\noalign{\smallskip}                                                                                                                                 
J0213$+$66  & 131.09 & $  4.66$ &  59.9 &  9.83 &  4.0*           & p  & --&--&--&--&--&  --&--&  --&c  & n  & 02 13 33.4                            &               \phantom{$-$}66 10 39   &  2.7   &  \ldots                   \\
     $ $    &        & $      $ &       &       &     \phantom{*} & a  & --&--&--&--&--&  --&--&  --&c  & n  & 02 13 11.5:                           &               \phantom{$-$}66 12 28:  &  2.4   &  \ldots                   \\
J0636$+$00  & 210.41 & $ -2.93$ &  34.1 &  9.56 &  5.1\phantom{*} & p  &  I& M& H&--& Z&  --& h&   N&c  & n  & 06 36 26.7                            &               \phantom{$-$}00 55 50   &  1.6   &  CGMW 1-0228              \\
     $ $    &        & $      $ &       &       &     \phantom{*} &a/c & --& M&--&--&--&  --&--&   N&-- & n  & 06 36 23.61                           &               \phantom{$-$}00 55 51.2 &  1.4   &  2MASX J06362361+0055513  \\
\multicolumn{12}{l}{$\cdots$} \\   
\end{longtable}
}

\twocolumn

\clearpage

\subsection{Results of counterpart search}

\begin{figure*}
\centering
\includegraphics[width=0.95\textwidth]{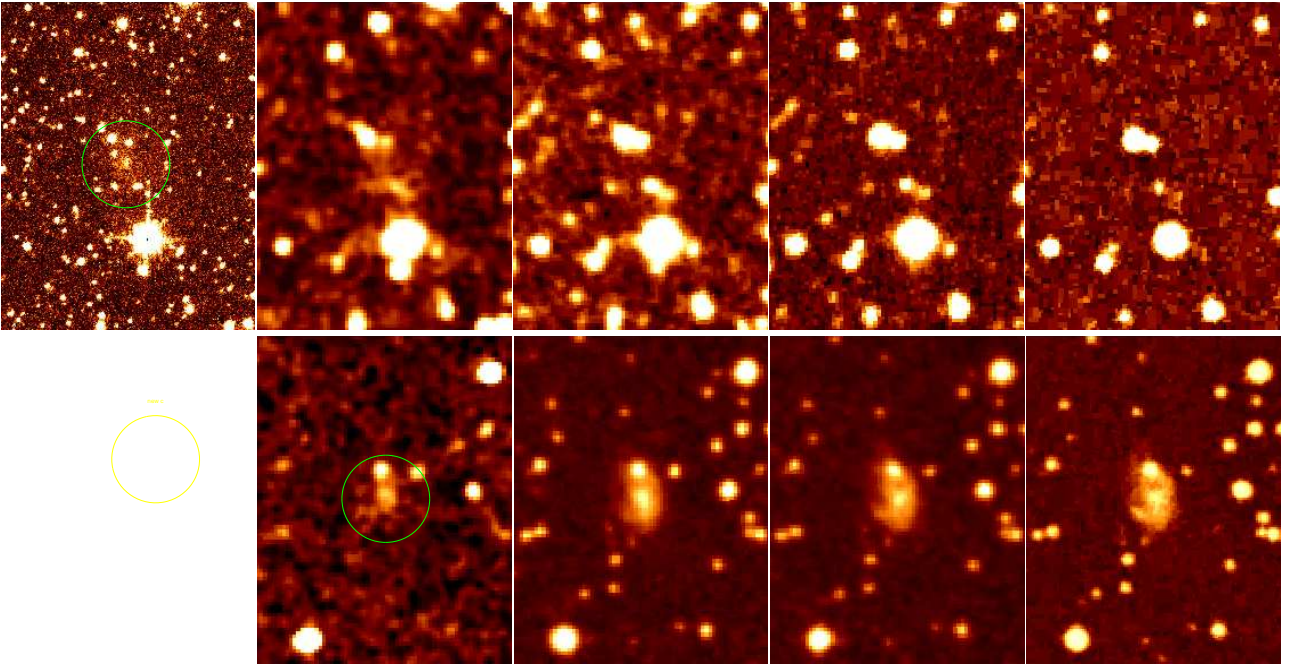}
\caption{Two examples of galaxies not previously catalogued in visual
  searches. From left to right: UKIDSS \K -band, 2MASS \K -band, DSS \II
  -band, DSS \R -band and SuperCOSMOS \B -band. Top: EZOA J2012+32 at an
  extinction of $A_B = 8\fm8$ is only visible in the deep \K -band
  image. Bottom: EZOA J2131+43 at a much lower extinction of $A_B = 1\fm3$
  is visible at all passbands (no deep \K -band image exists to-date).
}
\label{exampleplot}
\end{figure*}

For 141 of the 170 \HI\ detections we have found a cross-match (83\%). In
six cases (4\%) more than one counterpart contributes to the
\HI\ signal. For three \HI\ detections (2\%) more than one galaxy candidate
was found but no unambiguous counterpart could be decided on. The numbers
are comparable to the cross-match rates listed in \citet{staveley16} for
HIZOA-S. We have 67 new \HI\ detections, that is, the rate of new
\HI\ detections is only 39\% as compared to, \eg 67\% for HIZOA-S.  This is
understandable since the northern ZoA galaxies have been extensively
observed with the Nan\c{c}ay, Arecibo and Effelsberg radio telescopes in
targeted observations, and the northern ZoA is less severely affected by
selection bias due to extinction than the southern part. This is also
consistent with the fact that a fair fraction of the new detections, that
is, 25\% (or 19 detections) have no visible counterpart, which compares
well with the 22\% cited by \citet{staveley16} for HIZOA-S. On the other
hand, of our 103 previously known \HI\ detections only 12\% (or 12
detections) have no visible counterpart; they all had been serendipitously
detected in one of the less complete or less sensitive previous blind
\HI\ surveys. Finally, only 24 of the 67 previously \HI\ undetected
galaxies are published in a known galaxy catalogue.

Of the 147 cross-matched galaxies (where we count all of
Table~\ref{tabcross}a and~\ref{tabcross}b, and only the first entry each in
Table~\ref{tabcross}c), 99 (67\%) are listed in NED or else in SIMBAD
($N_S=3$). That means that 48 (33\%) cross-matched galaxies are new, or in
other words, they are not listed in the literature (at a wavelength other
than \HI ) but were found by us on at least one of the searched
images\footnote{They can be identified in Table~3 has having neither a name
  in Col.~13 nor an entry in the Col.~8 as having an \HI\ counterpart
  (labelled as `h').}. Figure~\ref{exampleplot} shows two examples: EZOA
J2012+32 is a typical example of a galaxy at high extinctions, whereas EZOA
J2131+43 is also visible at the optical passbands and was either missed in
optical searches or the particular area was not covered.

There are 39 (27\%) IRAS counterparts (plus one uncertain cross-match), 83
(56\%) 2MASS counterparts, and 23 (16\%) cross-matches have optical
velocity measurements listed in the literature.

The median distance between the \HI\ position and the actual counterpart
(using only definite candidates and non-confused profiles, $N=120$) is
$1\farcm2$, and 95\% of the cross-matches are found within $3\farcm3$ of
the \HI\ position. This compares well with the findings in
\citet{staveley16}, taking into account the difference in positional
uncertainties and beam sizes and the smaller sample size.


\section{Completeness and reliability }   \label{compl}

\begin{figure}
\centering
\includegraphics[width=0.45\textwidth]{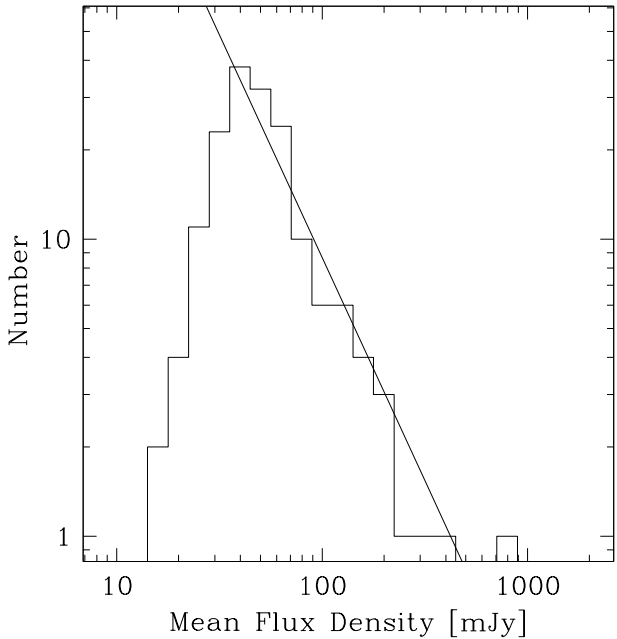}
\caption{Log of the mean \HI\ flux density $S$ (flux integral divided by
  the line width $w_{20}$) for all 170 EZOA detections. The black line
  corresponds to a slope of $-3/2$ that would be expected for a homogeneous
  distribution.
}
\label{histlfluxplot}
\end{figure}

Since our catalogue is of moderate size, we have determined a completeness
limit for the survey as defined in \citet{donley05} for the northern
extension of the HIZOA survey (HIZOA-N; $N$ = 77).
Figure~\ref{histlfluxplot} shows a histogram of the mean flux density $S$,
that is, the flux integral divided by line width $w_{20}$. Where $w_{20}$
was not available we added an offset of $24.5$\,\kms\ (as used in the error
calculation, see Sec.~\ref{sample}) to $w_{50}$. A nominal Euclidean power
law, $N(S) \propto S^{-2.5}_{\rm mean}$, laid by eye over the plot, shows
clearly that the completeness limit must lie between 40 and 50 mJy (given
the high scatter from the low numbers involved, together with the practical
difficulties of fitting with substantial uncertainties in both axes, we
considered the by-eye-fit to be sufficiently accurate). This limit
coincides well with the median mean flux density of 47\,mJy.  Assuming a
linear relationship between the peak flux and mean flux density, this
corresponds to a median peak flux of 66\,mJy, which is about $3\sigma$
above the nominal rms noise of the survey, 23\,mJy.

\subsection{Automated searches }

With the upcoming large \HI\ surveys to be conducted by the various SKA
precursors as well as SKA itself, visual searches of the large and numerous
data cubes are not feasible anymore. Various automated source detectors
exist (\eg \citealt{whiting12}, \citealt{serra12b}, \citealt{jurek12}),
some of which are made available in the software package
\sof\ \citep{serra15}. These have been extensively tested with
simulations. However, real data deviates from simulated data mainly in
non-uniform noise distribution, effects caused by RFI, the presence of
continuum sources etc. In particular continuum sources, causing variations
in the baseline which the automated source finders have difficulties
dealing with, are a major problem in the ZoA (\cf Fig.~7 in
\citealt{staveley16}). We were able to make some comparisons nonetheless.

\subsubsection{Preliminary EBHIS shallow source catalogue }

The preliminary EBHIS shallow source catalogue (described in
\citealt{floer15}\footnote{http://hss.ulb.uni-bonn.de/2015/4227/4227.htm})
was derived from a combination of an automated source finder, based on
wavelet de-noising (also included in \sof , \citealt{serra15}), with an
artificial neural network that determines whether a detected candidate is
likely to be a real source. Details on the method can be found in
\citealt{floer14}, \citealt{serra15} and \citealt{floer12}. We describe the
details of the comparison with our visual search in App.~\ref{prelim} since
the neural network discriminator needs more balanced training systems to
produce reliable results. Despite the caveats, a comparison is very useful
since sources detected in one but not both searches give valuable feedback
on the limitations of the respective search methods. In particular, we
checked sources from the EBHIS shallow source catalogue that were not found
in the visual searches. This approach added eventually three low-SNR
detections ($<6$) as well as one significant detection, EZOA J0637+03,
which, however, lies on a strongly varying baseline and could only be
confirmed through a cross-match with the HIZOA survey. No obvious source
was missed which proves the thoroughness of visual searches.


\subsubsection{The SOFIA package }

We have also used the automated source finders in the software package
\sof\ \citep{serra15} to compare with our visual search, taking into
account that \sof\ is still very much {\it under construction} and is
affected by various software issues currently addressed (note that most of
our testing was conducted on version 0.5).

\begin{table*}
\centering
\caption{Comparison of median rms and signal-to-noise ratio for different HI surveys} \label{rmstab}
\begin{tabular}{lcccc}
\hline
 & EBHIS & HIZOA-S & HIPASS-S & HIPASS-N \\  
\hline
\multicolumn{5}{l}{rms:} \\
overall (average)       & 23 mJy              & 6 mJy             & 13 mJy             & 14 mJy              \\
around detections      &  10.4 mJy ($N=170$) & 3.4 mJy ($N=883$) & 7.3 mJy ($N=4315$) &  7.7 mJy ($N=1002$) \\
matches with HIPASS     &  10.1 mJy ($N=37$)   & ---               & \multicolumn{2}{c}{8.5 mJy ($N=37$) }    \\
non-matches with HIPASS &   9.1 mJy ($N=14$)   & ---               & \multicolumn{2}{c}{---}                  \\
rms$_{\rm \,detected}$ : rms$_{\rm \,overall}$ &  0.45  & 0.57 & 0.56 & 0.55 \\
\noalign{\medskip}
\multicolumn{5}{l}{SNR:} \\
detections              &  7.2 ($N=170$) & 10.4 ($N=883$) & 8.5 ($N=4315$) &  8.0 ($N=1002$) \\
matches with HIPASS      &  8.0 ($N=37$)  & ---            & \multicolumn{2}{c}{9.1 mJy ($N=37$) }    \\
non-matches with HIPASS  &  6.1 ($N=14$)  & ---            & \multicolumn{2}{c}{---}                  \\
\hline
\end{tabular}
\end{table*}

\sof\ offers three different source finders in combination with three
different filters. We have tested these (excluding the 2D-1D wavelet filter
used for the aforementioned EBHIS source detection pipeline), with varying
input parameter settings, on a single
$12\degr\times12\degr$-cube\footnote{We also used a HIZOA Parkes multibeam
  data cube \citep{staveley16} as a test-set with similar results.}. We
found that none of the methods reliably detected sources, mainly due to
non-straight baselines. This limitation is basically due to the fact that
\sof\ has not been equipped with baseline subtraction which is obsolete for
the interferometric data cubes for which \sof\ is primarily
intended. However, the increased number of continuum sources in the ZoA
introduces many artefacts that cannot be fully removed and which affect
baselines also in interferometric data \citep{ramatsoku16}. In other words,
real data cubes from the ZoA differ considerably from the averaged
simulated interferometric cubes which were used to optimise the automated
source finders. Baseline subtraction, whether performed on the cube before
a source finder is used or during the search process, is a necessity for
any kind of ZoA data cubes and needs to be addressed.

\sof\ also performs source parametrisation using input catalogues; it
returned indeed most of our detections with improved positions and fitted
\HI\ parameters. Here we found that the recommended Smooth $+$ Clip finder
\citep{serra12b} in combination with the Noise Scaling filter was the best
choice by having the most complete return of input sources
\citep{popping12}.

\subsection{Comparison with other \HI\ surveys } \label{complit}

During our visual search, we used the more sensitive HIZOA catalogues to
confirm doubtful detections in the overlap region ($-5\degr \leq \delta
\leq +25\degr$) and to ensure that no source was overlooked.  Only four
sources were not found in our initial search, all with an SNR between 5 and
7 and all affected by varying baselines.

As a fully independent check, we compared our final list of detections with
HIPASS (\HI\ Parkes All Sky Survey), which is a blind \HI\ survey of the
full southern hemisphere plus a northern extension at an intermediate
sensitivity (rms $= 13-14$\,mJy), also conducted with the Parkes radio
telescope (\citealt{meyer04}; hereafter HIPASS-S; \citealt{wong06};
hereafter HIPASS-N). There are 39 detections in common: 27 from HIPASS-S
and 11 from HIPASS-N, with one further detection recorded in the bright
galaxy catalogue, which is based on an older version of the HIPASS-S
catalogue \citep{koribalski04}.

For the completeness check, we have looked at all HIPASS sources that we
have not detected. Most of these were indeed too faint for our survey, that
is, their peak flux density is $<50$\,mJy. The brightest, with a peak of
$\sim 100$\,mJy (HIPASS\,J1917+11), is near a continuum source and was thus
not detected by us. Four HIPASS sources with peak flux densities ranging
from $60 - 90$\,mJy are visible in the EBHIS cubes but they are not
distinctly distinguishable from noise. We conclude that our search did not
miss any significant detections.

For the reliability check, we looked at the 14 EZOA detections in the
overlap region with HIPASS which were not detected by HIPASS. We
cross-checked each position with the HIPASS data
archive\footnote{http://www.atnf.csiro.au/research/multibeam/release} and
find that five of the detections show strong baseline variations in HIPASS,
which likely affected the automated HIPASS source detection, seven are
visible but have clearly a low signal-to-noise ratio, and two show obvious
profiles (EZOA J0552+22 and EZOA J0640$-$01).


To explore why the sources were not detected with the more sensitive
HIPASS, we have compared the rms (after baseline fitting) around
\HI\ detections for the four surveys EZOA, HIZOA-S \citep{staveley16},
HIPASS-S and HIPASS-N, see Table~\ref{rmstab}.  As expected, the rms around
detected sources tends to be smaller than the average rms across a full
survey: \eg whereas the overall rms of the HIZOA-S survey is 6\,mJy, the
median rms around the detected sources is only 3.4\,mJy. If we calculate
the ratio of these `detected' rms values with the overall rms, we find that
the ratio is slightly lower for EZOA (0.45 versus 0.57, 0.56 and 0.55 for
the three other surveys, respectively).

We can also compare the SNR: While the median SNR for HIZOA-S is high at
10.4 (this survey used deliberately a high cut-off with the aim to be
reliable), the median SNR of EZOA at 7.2 is also lower than for HIPASS,
albeit marginally so (8.5 for HIPASS-S and 8.0 for HIPASS-N), as one would
expect for a visual versus automated search.

The 14 EZOA detections not detected by HIPASS have a slightly lower median
rms and also a lower median SNR than the average, indicating that they were
only found with EBHIS because they are in low-rms areas (for example,
overlap regions in the mosaiced survey).  Furthermore, 11 of these
detections are in the HIPASS-N region where the intrinsic rms is higher due
to the low elevation of the Parkes radio telescope (\cf discussion in
\citealt{wong06}).

\subsection{Source parameters }

We compared the extracted \HI\ parameters with published values. We found
reasonable overlap with HIPASS ($N=38$), HIZOA ($N=32$) and \nan\ radio
telescope measurements (NRT; $N=35$)
(\citealt{kk18,paturel03,theureau98b,chamaraux90,martin90}).  For the
comparison we excluded obviously confused cases.  Table~\ref{comptab} gives
the mean difference in the parameter, the median and the standard deviation
for the detections in common (\ie where the parameters were available; note
that for HIPASS we use their width-maximised $v_{50}$, $w_{50}$ and
$w_{20}$). No statistically significant systematic effects (\ie at the
$3\sigma$ level) are noticeable except for the marginal case of the flux
comparison with HIZOA which improves to a mean of $-13.3\pm5.9$ when only
sources with SNR $>6$ are used.

\begin{table}
\centering
\caption{Comparison of EZOA parameters} \label{comptab}   
{\small
\begin{tabular}{lrrrr}
\hline
Parameter [unit] & N & \multicolumn{1}{c}{Mean} & Median & Std dev \\  
\hline
a) HIPASS (PKS) & & & & \\
\noalign{\smallskip}
$v_{50}$ [\kms ]    & 36 &  $-1.3\pm2.2$ & $\phantom{}-1.1$ & 13.0 \\
$w_{50}$ [\kms ]    & 35 &  $-9.6\pm5.1$ & $\phantom{}-3.6$ & 29.9 \\
$w_{20}$ [\kms ]    & 30 & $-10.3\pm8.5$ & $\phantom{}-5.5$ & 46.5 \\
Flux integral [\%]  & 36 &  $-13.6\pm6.4$ & $\phantom{}-7.8$ & 38.2 \\  
\hline
b) various NRT & & & & \\
\noalign{\smallskip}
$v_{50}$ [\kms ]    & 35 & $0.1\pm1.3$ & $\phantom{}0.0$ &  7.8 \\
$w_{50}$ [\kms ]    & 32 & $6.1\pm4.3$ & $\phantom{}3.0$ & 24.4 \\
$w_{20}$ [\kms ]    & 32 & $1.3\pm5.3$ & $\phantom{}6.5$ & 29.8 \\
Flux integral [\%]  & 34 & $8.5\pm5.2$ & $\phantom{}10.5$ & 30.2 \\  
\hline
c) HIZOA (PKS) & & & & \\
\noalign{\smallskip}
$v_{50}$ [\kms ]    & 31 & $-2.3\pm1.7$ & $\phantom{0}0.0$ & 9.6 \\
$w_{50}$ [\kms ]    & 31 & $0.6\pm5.4$ & $\phantom{}10.0$ & 30.0 \\
$w_{20}$ [\kms ]    & 27 & $8.0\pm5.7$ & $\phantom{}15.0$ & 29.8 \\
Flux integral [\%]  & 31 & $-20.5\pm6.3$ & $\phantom{}-17.1$ & 35.3 \\  
\hline
\multicolumn{3}{l}{e) \sof\ (for $w_{50}<200$\,\kms )} & & \\
\noalign{\smallskip}
$v_{50}$ [\kms ]    & 109 & $-0.6\pm1.6$ & $\phantom{}-3.0$ & 16.8 \\
$w_{50}$ [\kms ]    & 109 & $8.2\pm2.7$ & $\phantom{}8.0$ & 28.6 \\
$w_{20}$ [\kms ]    &  97 & $6.2\pm3.1$ & $\phantom{}11.0$ & 30.2 \\
Flux integral [\%]  & 109 & $-4.3\pm8.6$ & $\phantom{0}6.8$ & 90.0 \\  
\hline
\end{tabular}
}
\end{table}

\subsubsection{Automated source finders }

The EBHIS preliminary shallow source catalogue lists all
\HI\ parameters. They compare well with our parameters except for an offset
in \wtw\ which is due to the fact that we used hanning smoothed data; since
the catalogue is only preliminary we will not go further into details.

Since we were able to extract most of the sources found in the visual
search with \sof\ we investigated the source parameters from the most
successful run using the Smooth $+$ Clip finder and the Noise Scaling
filter. Though the velocities compare well, the line width measurements are
highly affected by the size allocated by \sof\ for parametrisation being
restricted to 22 pixel (that is, in the case of EBHIS only about 225 \kms
). This feature is being addressed by the \sof\ developers. Hence, for the
comparison we used only detections with $w_{50}<200$\,\kms , see the last
entry in Table~\ref{comptab}. The parameters now compare well, only
\wfi\ shows a $3\sigma$ offset.

Since \HI\ positions of single dish observations have large uncertainties
due to the large beam size, we wanted to know if the accuracy of the source
position varies depending on the parametrisation algorithm. For the
comparison, we have used the positions of the optical/NIR counterparts
(Sec.~\ref{cmatches}). While the positions read off visually from the cubes
are understandably less accurate (these positions were usually chosen as
the ($x$,$y$)-pixel where the profile has its peak), the positions from the
\mir\ and \sof\ fits as well as the positions in the \lar\ catalogue are of
comparable precision. Table~\ref{compdisttab} lists the mean, median and
standard deviation of the distribution in distance between the counterpart
and \HI\ position (in arcminutes) for definite candidates and non-confused
profiles ($N=120$).

\begin{table}
\centering
\caption{Comparison of \HI\ position accuracy using optical and NIR
  cross-matches} \label{compdisttab}
{\small
\begin{tabular}{lrlll}
\hline
Method & N & Mean & Median & Std dev \\  
\hline
Visual & 120 & $ 1\farcm92\pm0\farcm09 $ & $1\farcm75$ & $1\farcm01$   \\
\mir\  & 120 & $ 1\farcm54\pm0\farcm11 $ & $1\farcm22$ & $1\farcm22$   \\
\sof\  & 117 & $ 1\farcm46\pm0\farcm09 $ & $1\farcm24$ & $0\farcm97$   \\
\lar\  &  64 & $ 1\farcm49\pm0\farcm14 $ & $1\farcm18$ & $1\farcm11$  \\
\hline
\end{tabular}
}
\end{table}

These comparisons give an idea on how the parameters may vary when a
different software is used, whereas a comparison with the literature points
out differences due to instrument and observing strategies. 
Table~\ref{comptab} shows that there are no significant differences (beyond
the one exception mentioned above) and we conclude that variations in the
data (with telescope and instrument settings) do not increase the
uncertainties in the parameters.

\section{HI properties of the sample }   \label{hiprop}

\begin{figure}
\centering
\includegraphics[height=0.92\textheight]{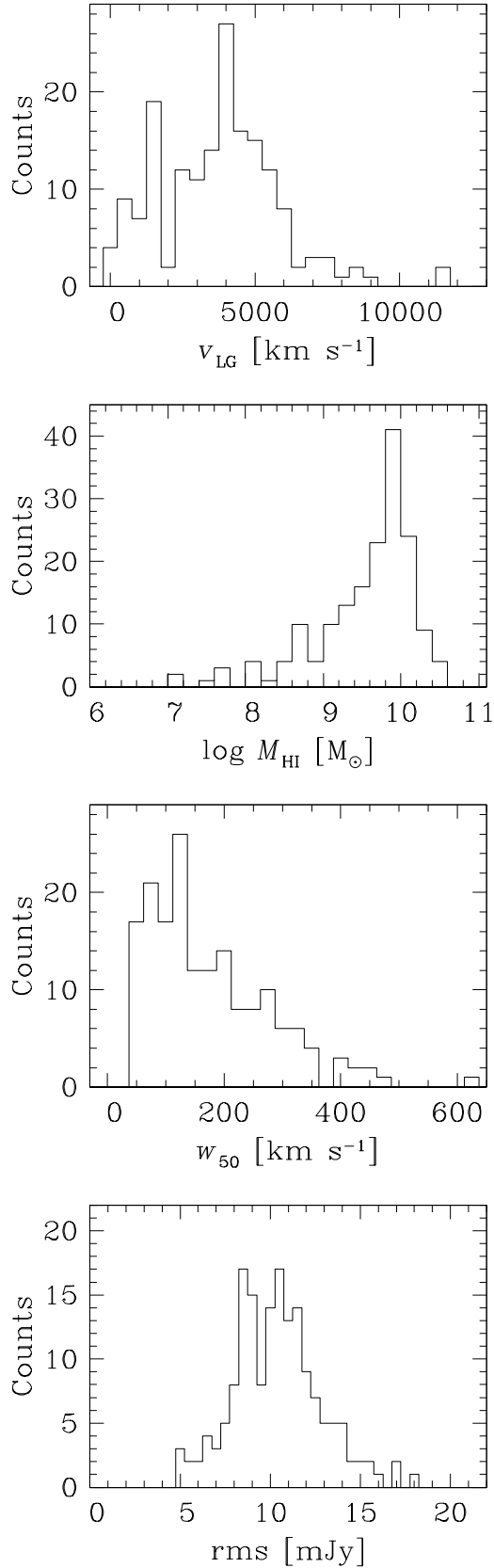}
\caption{HI\ parameters of the 170 detections in the EZOA survey. From top
  to bottom the histograms display the radial velocity $v_{\rm LG}$, the
  \HI-mass distribution, the line width $w_{50}$, and the clipped rms noise
  at the position of the detected galaxy.
}
\label{histparam}
\end{figure}

To characterise the \HI\ properties of our sample, we present some of the
parameters in Fig.~\ref{histparam}. The top panel shows the recessional
velocities in the Local Group frame of reference. Though the survey's
sensitivity drops off quickly for $ \approxgt \,5000$ \kms , there is a
pronounced peak at 4000 \kms\ which is clearly due to large scale
structures (see discussion in Sec.~\ref{lss}). The next panel shows the
distribution of \HI\ masses, ranging from \lmhi\ = 7.0 (for EZOA J0630+08
at $v_{\rm LG} = 250$ \kms , no counterpart visible but detected by HIPASS
and HIZOA) to 10.6 (for EZOA J2204+48 at $v_{\rm LG} = 11,66$ \kms\ with an
edge-on visible galaxy not recorded in the literature as counterpart; the
SNR, however, is only 6.7). The mean is 9.6 and the median is 9.8.


\begin{figure}[th]
\centering
\includegraphics[width=0.44\textwidth]{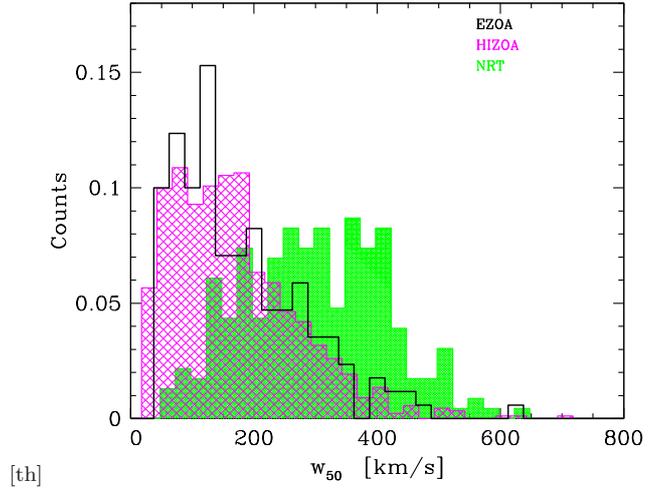}
\caption{\HI\ line width distribution: the two blind surveys EZOA (black
  open histogram) and HIZOA (hashed magenta) are compared to a NIR-selected
  survey (green filled histogram) conducted with the NRT \citep{kk18}.
}
\label{histw50comp}
\end{figure}

\begin{figure}[th]
\centering
\includegraphics[width=0.44\textwidth]{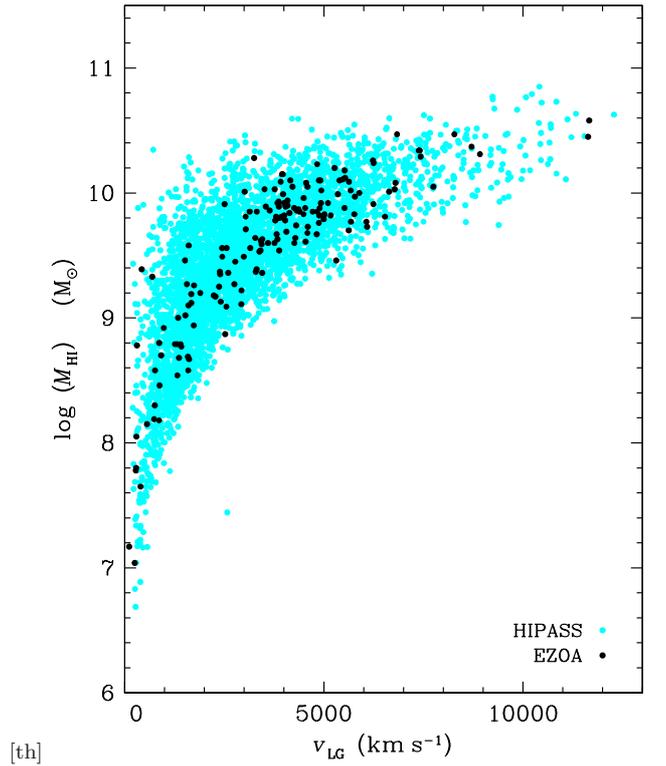}
\caption{\HI\ mass versus velocity in the Local Group frame for the EZOA
  (black) and HIPASS samples (light blue; \citealt{meyer04}). 
}
\label{distmhi}
\end{figure}

The distribution of line widths measured at half peak, $w_{50}$, is highly
skewed with a peak at $\sim\!100$\,\kms , a mean of 176\,\kms\ and a median
of 146\,\kms , very similar to the HIZOA-S survey (with a mean and median
of 163\,\kms\ and 147\,\kms , respectively). It is interesting to note that
the line width distribution of a pointed survey of NIR-selected targets,
as, for example, presented by \citep{kk18}, shows a more Gaussian
distribution with a higher mean of 288 \kms\ that is the same as the
median, as shown in Fig.~\ref{histw50comp} as the filled green
histogram. It is obvious that many galaxies with narrow line widths (often
low-surface brightness dwarf galaxies) are missed in the pointed
observations. This is likely due to the high Galactic foreground extinction
and the low sensitivity of the NIR to late type spirals and irregular
galaxies which means they rarely appear in NIR-selected samples in the ZoA.

The iteratively-clipped rms noise around each detection is shown in the
bottom panel. Whereas the overall rms of the EBHIS survey is 23 mJy, the
rms around the detections, as discussed in Sec.~\ref{complit}, is
considerably lower, partly due to the clipping but also because the overall
rms refers to the full velocity range ($-600 < cz < 18000$\,\kms ) while most
of our detections are below 6000 \kms . The histogram also shows a
considerable contribution from very low-rms detections most of which are
located in the overlap regions between the mosaiced scans.

Figure~\ref{distmhi} shows the distribution of \HI\ mass as a function of
velocity in the Local Group frame for the EZOA and HIPASS-S detections. The
low statistics of the EZOA sample makes a comparison difficult, but the
overall shape of the distributions are comparable and no undue outliers are
apparent.

Two notable detections are the one with the lowest \HI\ mass and the one
with the largest line width. The lowest \HI\ mass (\lmhi = 7.0) was
measured for EZOA J0629+08 at $v_{\rm LG} = 250$ \kms\ for which no
counterpart could be found; it was detected, however, in the HIPASS and
HIZOA surveys.  The largest line width (\wfi = 616 \kms ) belongs to a
marginal detection (EZOA J0358+54) with no visible counterpart (it has a
high extinction of $A_B = 6\fm7$).

\section{Large-Scale Structure Analysis }  \label{lss}

Despite the low sensitivity of the shallow EBHIS survey, 39\% of the
\HI\ detections are new ($N=67$), and only five of these have previously
recorded optical redshifts. Twenty-six of the new \HI\ detections have a
visible counterpart {\it not} previously recorded in the literature, and a
further 17 have no visible counterpart so far (likely due to a mix of
difficult to find low-surface brightness galaxies and the higher extinction
in the survey region). In addition, 34 of those detections which already
have been recorded in existing \HI\ catalogues have also no optical or NIR
counterpart in the literature, though in 22 of these cases we found such a
counterpart on at least one of the images we searched. Though the numbers
of new galaxies are small, they show the power of systematic blind
\HI\ surveys in the ZoA.


\subsection{Nearby galaxies}

Fifteen of our detections are found within 11\,Mpc, six of which belong to
the Maffei group \citep{karachentsev05}. Three detections were not observed
in \HI\ before, and four have no optical or NIR counterpart. The 15
detections are dominated by dwarf galaxies: the \HI\ masses range from
\lmhi = 7.0 to 9.4, with a median of 8.1. The line widths are also
considerably smaller, with a median of $w_{50} = 94$\,\kms\ as compared to
146\,\kms\ for the full catalogue.

No previously unidentified galaxy was found in this volume, proving that
there is no hidden massive galaxy in the northern sky that would
significantly influence the motion of the Local Group with respect to the
microwave background.

\subsubsection{Closest detection: IC 10} \label{ic10}  

The closest detection is that of IC\,10 (EZOA J0020+59) at \vhi $=
-346$\,\kms ; the velocity with respect to the Local Group is $-84$\,\kms
. We adopt 0.74\,Mpc for its distance, as determined by \citet{tully13}
based on measurements of Cepheids and the tip of the red giant branch. Our
velocity agrees well with the range found in the literature ($\sim -352 -
-342$\,\kms ). The \HI\ emission covers at least 23 frames in the cube, and
it is likely that a more meticulous extraction of the profile would improve
the parameters.

\subsubsection{HIZSS\,003 }  \label{ss3}

The second closest detection in our catalogue (EZOA J0700$-$04) is
HIZSS\,003 which was only recently detected in the Parkes HIZOA Shallow
Survey \citep{henning00a}.  \citet{begum05} observed this galaxy with the
VLA and found that the \HI\ emission comes from two dwarf galaxies. Both
are faintly visible on the deep NIR images; we give their coordinates in
our cross-match table.  Together with EZOA J0630+08 it is the least massive
dwarf in our catalogue (\lmhi = 7.1 and 7.0, respectively).

\subsubsection{Maffei group}

The Maffei group (\eg \citealt{karachentsev05}) consists of the IC\,342
subgroup and the Maffei 1 subgroup; only the Maffei 1 subgroup lies deep in
the plane and was thus detected by us. Of the 8 known members in the Maffei
1 subgroup, we have detected six: Dwingeloo 1 and 2, KK 11 and 12, Maffei 2
and MB 3.  We did not detect MB 1 (\vhi $= 59$\kms ,
\citealt{huchtmeier03}) since it is faint and close to Dwingeloo 1 (as well
as to the Galactic \HI ). Maffei 1 is an S0 galaxy and has not been
detected in \HI . A ninth, possible member (KKH\,6) lies outside our search
area. No new member was uncovered.

\subsubsection{New \HI\ detections at D < 11\,Mpc}

Of particular interest is EZOA J2120+45 or 2MASX J21204618+4516221 which
lies in the Local Volume at a distance of 7.5\,Mpc and was not detected in
\HI\ before. Its \HI\ mass is $\log M_{\rm HI} = 8.2$. In addition, EZOA
J0506+31 and EZOA J0301+56, at the edge of the local volume at 10.1\,Mpc
are newly found galaxies. Both are dwarf galaxies at $\log M_{\rm HI} =
8.3$ and 8.6, respectively. EZOA J0301+56 does not have an optical
counterpart.

\subsection{Large-scale structures }

\begin{figure*}
\centering
\includegraphics[width=0.9\textwidth]{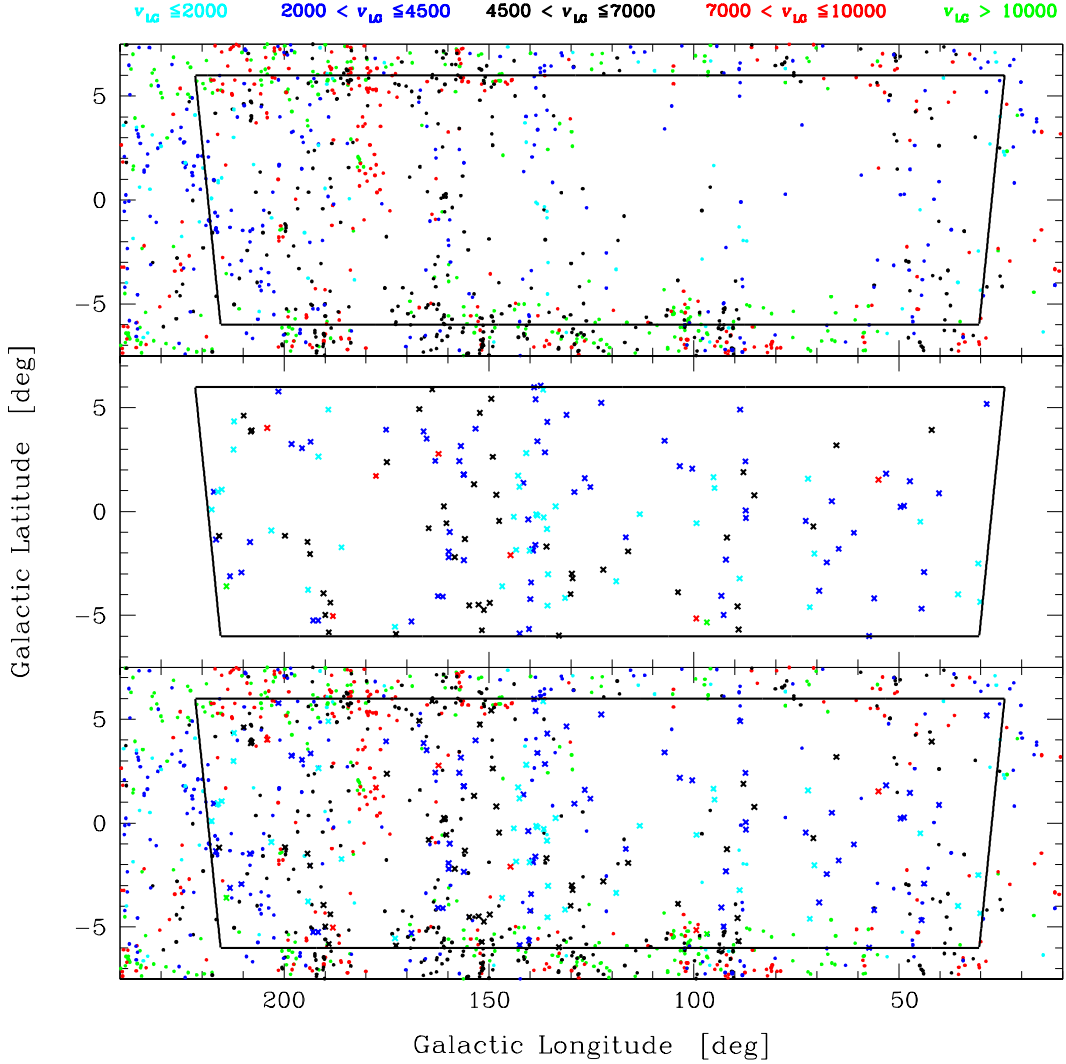}
\caption{Distribution of galaxies with velocity measurements in Galactic
  coordinates. Top panel: distribution of galaxies available through
  HyperLEDA; middle panel: distribution of the 170 \HI-detected galaxies;
  bottom panel: both distributions are shown together. The survey area is
  indicated by the dashed line ($\delta\leq -5\degr$, $|b| < 6\degr$). The
  symbols are colour-coded as a function of velocity. Note the predominance
  of galaxies around $\ell \sim 150\degr$ and $\ell \sim 90\degr$.
}
\label{mapplot}
\end{figure*}

\begin{figure*}
\centering
\includegraphics[width=0.5\textwidth,angle=270]{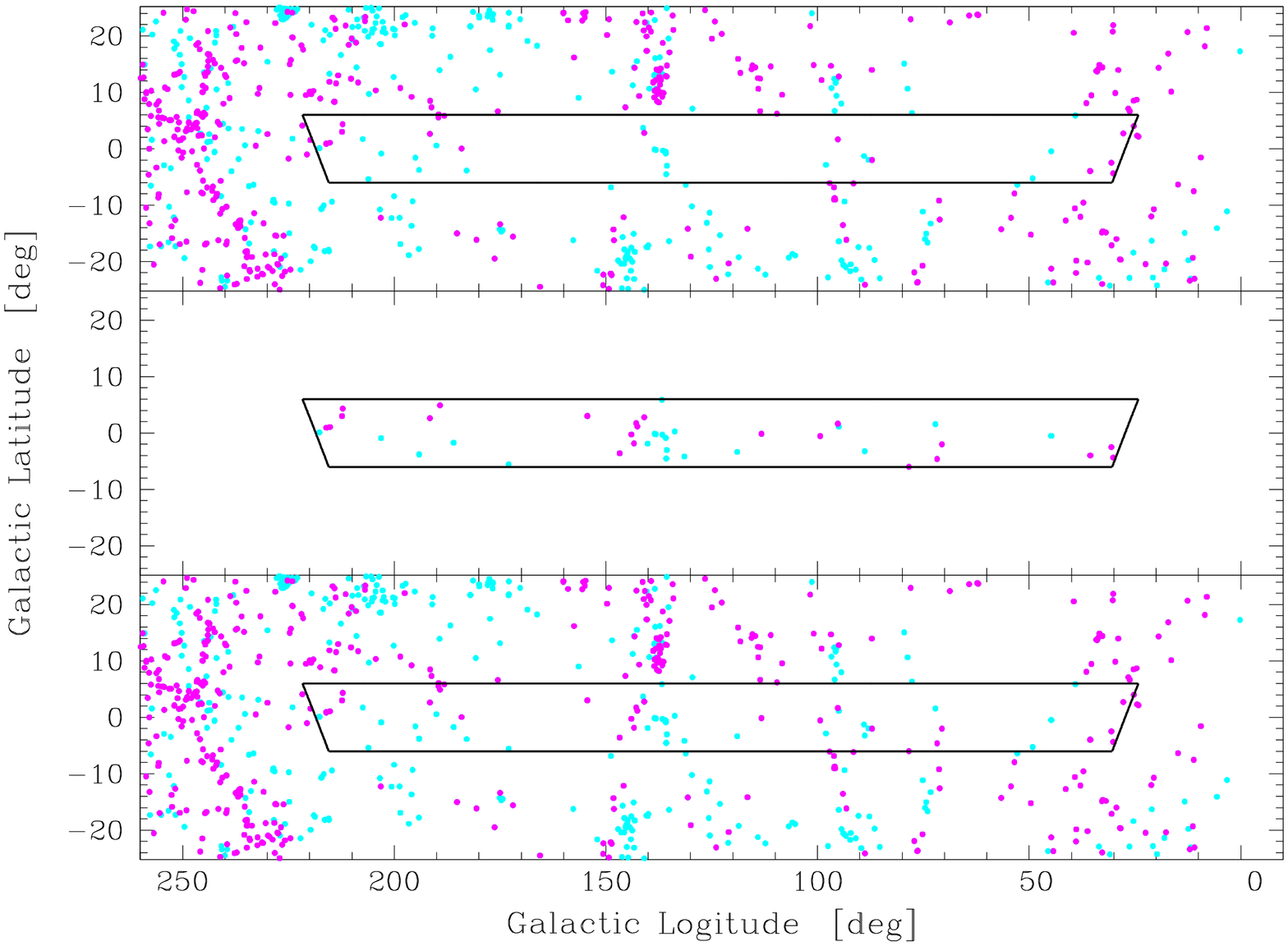}
\caption{Same as Fig.~\ref{mapplot} for nearby galaxies: light blue: \vlg <
  1000\,\kms ; magenta: 1000\,\kms < \vlg < 2000\,\kms .
}
\label{map15plot}
\end{figure*}

\begin{figure*}
\centering
\includegraphics[width=0.5\textwidth,angle=270]{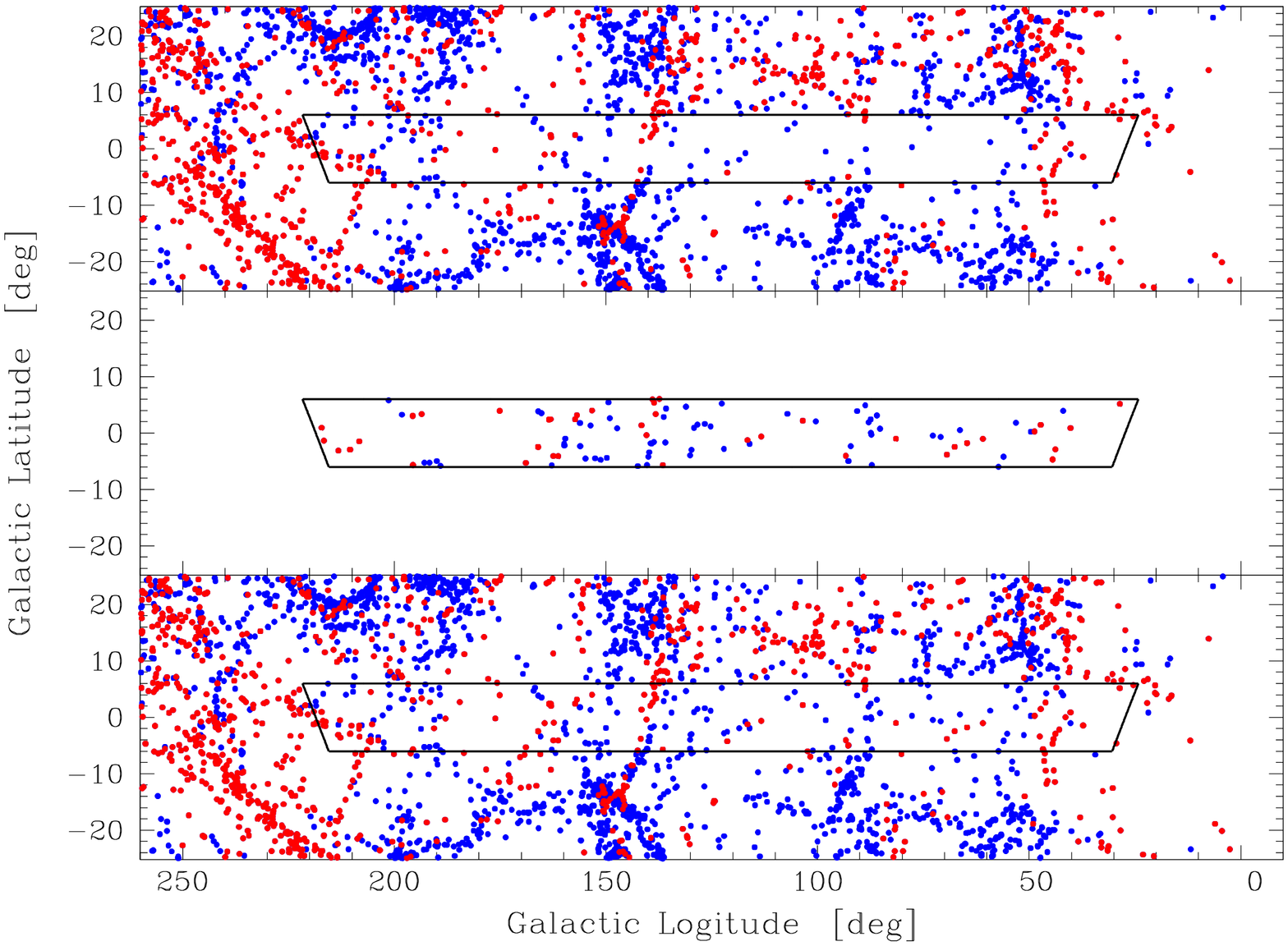}
\caption{Same as Fig.~\ref{mapplot} for galaxies at intermediate
  velocities: red: 2000\,\kms < \vlg < 3500\,\kms ; blue: 3500\,\kms < \vlg
  < 5000\,\kms .
}
\label{map40plot}
\end{figure*}

Figure~\ref{mapplot} shows the distribution of the 170 \HI\ detections in
Galactic coordinates colour-coded by velocity (middle panel). For interest,
we also show the velocity measurements available in the literature, using
HyperLEDA (status March 2019, using {\tt objtype} = `G'), top panel, and
our new detections added in, bottom panel. Two features are immediately
obvious: the Supergalactic plane (SGP, at $\ell \sim\!140\degr$) has been
more firmly established, and the empty patch at around $60\degr < \ell <
80\degr$ and $-5\degr < b < +1\degr$ has been filled. We discuss these and
other features in detail below, using plots of individual redshift slices
that cover a larger area ($260\degr > \ell > 24\degr, |b| \leq 25\degr$;
Figs~\ref{map15plot} and~\ref{map40plot}) where individual large scale
structures can be traced more easily.

Figure~\ref{map15plot} shows the nearby galaxies up to 2000\,\kms .  Though
the number of detections in this slice is small ($N=40$), both
aforementioned features are present:

(1) At $140\degr<l<150\degr$, the connection of the Supergalactic Plane
across the ZoA shows six new galaxies in the narrow velocity range
1350\,\kms < v < 1750\,\kms, three of which have no optical
counterpart. Just above our search area, at $b\sim 12\degr$, are three
known galaxy groups at similar velocities (HDC 248, \citealt{crook07},
[TSK2008] 0139 and [TSK2008] 0141, \citealt{tully08}). The new galaxies
seem to be part of a filament connecting these groups into and possibly
across the ZoA towards the south.


(2) The previously empty patch around $l\sim70\degr$ shows three new
galaxies with $980 <$ v $< 1530$\,\kms . It looks like a continuation of a
filament below the ZoA but a deeper survey is required to strengthen this
connection.

Figure~\ref{map40plot} shows galaxies between 2000 and 5000\,\kms ; there
are 92 detections in total. The two features mentioned above are also
prominent in this slice, but there are other noticeable structures as well.

(3) The innermost ZoA gap of the SGP at $l \sim 140\degr$, that is, between
$b=0\degr$ and $b=-2\degr$, is now filled in both redshift bins.

(4) The previously empty patch around $l\sim70\degr$ shows seven
detections, with four having redshifts in the very narrow redshift range
$3380-3460$\,\kms , possibly indicating a new filament crossing the ZoA
from South-East to North-West -- again, more data is required to make a
definite statement. One of these galaxies was also detected by HIPASS (it
does not appear in the HyperLEDA sample since it was labelled `HI' and not
as a galaxy, having no published optical counterpart). All seven galaxies
are visible in the NIR.

(5) The tail-end of the Perseus--Pisces filament that crosses the ZoA in
Cygnus at $l \sim 90\degr$ and which has been discussed extensively by
\citet{ramatsoku16} and \citet{kk18}, has also been strengthened despite
the drop in sensitivity of our survey for redshifts above 4000\,\kms .

(6) In the area of $150\degr < \ell < 170\degr$, two filaments at different
redshifts, seemingly crossing each other, have obtained more galaxies: the
lower-redshift filament crosses from South-East to North-West and seems to
go parallel to the SGP. The higher-redshift filament crosses from
South-West to North-East, that is, from the Perseus cluster
($v\sim5400$\,\kms ) towards the intermediate-rich galaxy cluster
ZwCl\,0731.9+3125 ($v\sim 4550$\,\kms ). It is possibly another branch of
the Perseus-Pisces filament, curving back towards lower velocities, whereas
the other branch continues strictly North and outward, that is, through
$\ell = 160\degr$ ($v\sim6000$\,\kms ) to Abell 569 ($v\sim5900$\,\kms )
\citep{kk18}.

(7) Another noticeable structure in our plot, though we have not added any
new data here, is a tenuous filament at around $v\sim3000$\,\kms , crossing
the ZoA at $40\degr < \ell < 50\degr$ in a North--South direction.

While the Perseus--Pisces supercluster is one of the largest and most
prominent large-scale structures in the northern sky, it lies just outside
the sensitivity of our survey ($v>5000$\,\kms ). We will, however, be able
to trace it with the upcoming full EBHIS survey.

\section{Summary and conclusions}   \label{concl}

We used the first pass of the EBHIS survey (with a sensitivity of
23\,mJy\,beam$^{-1}$ at a velocity resolution of 10.24\,\kms ) to extract a
catalogue of galaxies in the northern ZoA ($\delta \geq -5\degr$ and $|b| <
6\degr$).  We have found 170 detections in \HI , 67 of which are new
detections and only 24 of these were previously recorded in the literature
as (optical) galaxies. These numbers demonstrate the power of blind
\HI\ surveys in searching for galaxies in the ZoA, and even more so since
the here-presented survey is only a very shallow one.

The EZOA \HI\ parameters and positions are of good quality: they compare
well with the literature and other source parametrising algorithms.  The
positional $1\sigma$ uncertainty is found to be $1\farcm2$, with 95\% of
the cross-matches lying within $3\farcm3$ of the \HI\ position.

Blind \HI\ surveys and pointed surveys of optical or NIR selected targets
in the ZoA differ significantly in the \HI\ line width
distribution. Whereas the pointed observations show nearly a Gaussian
distribution with a mean of $\wfi \sim\!300$\,\kms , blind surveys find
many more narrow-line-width detections which often come from low-surface
brightness dwarf galaxies that are rarely visible at higher Galactic
extinction levels and are thus missed in pointed observations of an optical
or NIR selected sample. Thus, we have found two new dwarf galaxies located
at the edge of the Local Volume at 10.1\,Mpc.

With 62 new redshift measurements in the 2280 square-degree northern ZoA
strip, we find that most prominent large-scale structures crossing this
strip have been established more firmly. We also found new galaxies in a
previously empty region around 70\degr\ in Galactic longitude and slightly
below the Galactic plane. The full EBHIS survey, which will be available in
2020 and will have a sensitivity comparable to the HIPASS survey in the
South, will be most valuable for the ZoA research and the full-sky cosmic
flow analyses by filling in the still persistent gap in the northern ZoA.

\section*{Acknowledgements}

This research has made use of: the HyperLEDA database; the NASA/IPAC
Extragalactic Database (NED) which is operated by the Jet Propulsion
Laboratory, California Institute of Technology, under contract with the
National Aeronautics and Space Administration; the SIMBAD database,
operated at CDS, Strasbourg, France; the SuperCOSMOS Sky Surveys; the WFCAM
and VISTA Science Archives, operated at the Royal Observatory of Edinburgh
(WFAU); the Sloan Digital Sky Survey which is managed by the Astrophysical
Research Consortium for the Participating Institutions. ACS thanks the
South African NRF for their financial support.




\bibliographystyle{mn2e} 
\bibliography{ZoA_bibfile} 




\newpage
\appendix

\section{The preliminary EBHIS shallow source catalogue }  \label{prelim}

The EBHIS shallow source catalogue comprises 89 sources\footnote{One
  additional source with $b=6\fdg15$ has also a cross-match in the EZOA
  catalogue (EZOA J0314+64) but since the detection goes beyond the edge of
  our (cut-out) data cube, the fitted parameters are uncertain and thus
  this detection is excluded from this analysis.} for $|b|\le 6\degr$. A
comparison with our catalogue confirms 73 of these sources as real, the
others are caused either by a combination of high noise peaks and baseline
variation or are too faint for a decision. Among the EBHIS {\it candidates}
(\ie detections found by the automated source finder before being
classified by the neural network algorithm; $N=$\,26,586 for $|b|\le
6\degr$) are a further 48 real detections (0.2\%).

\begin{figure}
\centering
\includegraphics[width=0.45\textwidth]{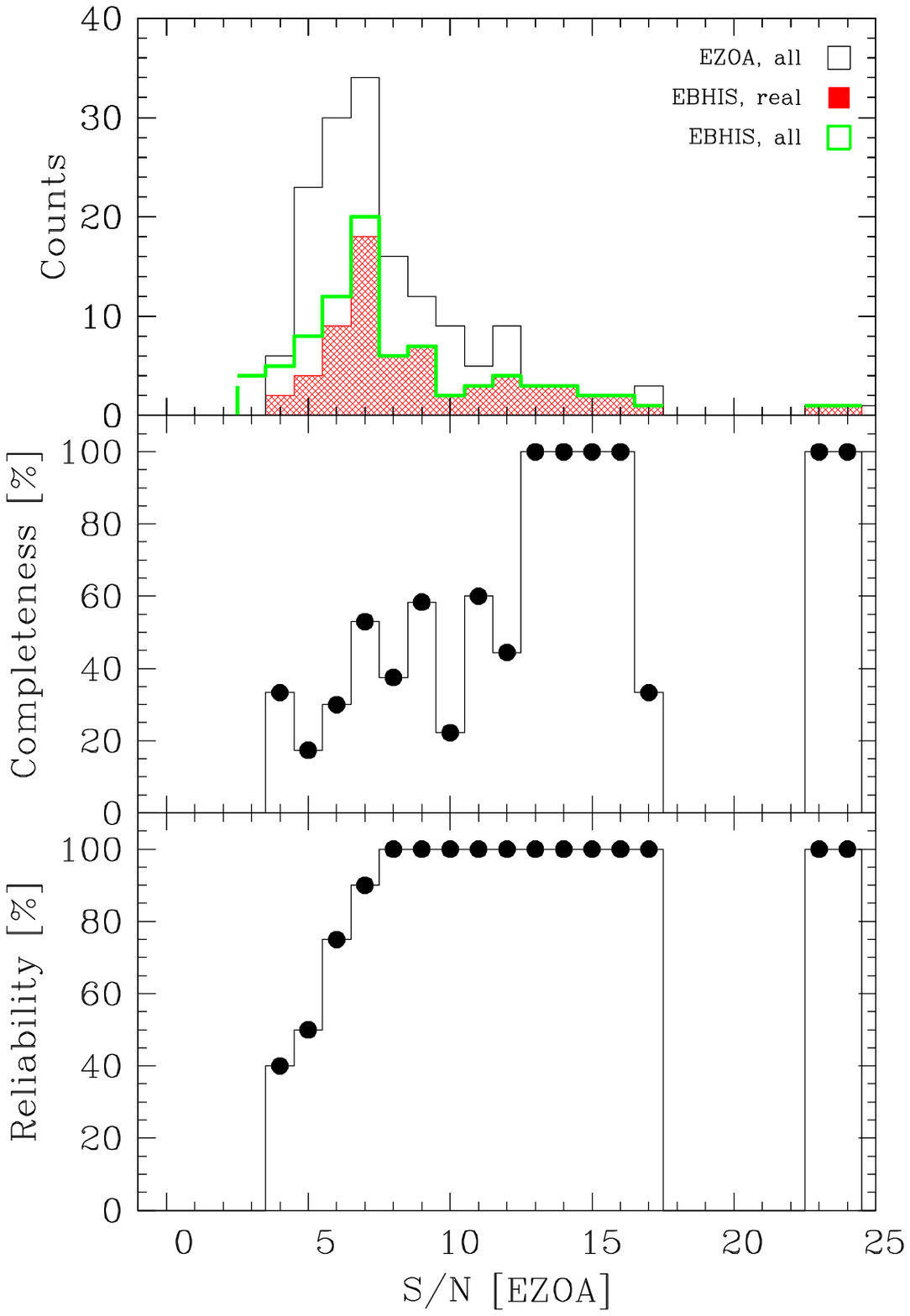}
\caption{Top panel: Histogram of SNR of all EZOA detections (black), all
  EBHIS source candidates (thick green), and real EBHIS sources (hashed
  red). Middle panel: Completeness of the EBHIS shallow source catalogue as
  a function of SNR. Bottom panel: Reliability of the EBHIS sources as a
  function of SNR.
}
\label{complrelplot}
\end{figure}

Taking our EZOA catalogue as reference, the EBHIS shallow source catalogue
is complete only for sources with a peak signal-to-noise ratio (SNR) $>13$
but is reliable from SNR = 8 onwards, as shown in Fig.~\ref{complrelplot},
middle and bottom panel, respectively. This is only an estimate, though,
since the statistics are low, as shown in the top panel of
Fig.~\ref{complrelplot}.

The overall reliability of the EBHIS shallow source catalogue in the ZoA is
82\%, improving to 98\% for flux densities above $10\,\Jykms$.  The rather
low reliability is likely due to the influence of the large number of
continuum sources in the Galactic plane on the baselines \citep{staveley16}
and thus on the automated source finding and classification. In addition,
the low number statistics combined with large uncertainties add a large
error to the rates we have found.

Note that during the fine-tuning of the pipeline used for the compilation
of the EBHIS shallow source catalogue, a comparison with our visual search
was used to improve the neural network algorithm that discriminates between
good and unreliable candidates. A comparison of this initial catalogue with
the latter one confirms that, as intended, the latter catalogue is somewhat
more reliable but less complete; this concerns mainly detections in the
range $7\, \approxlt \, {\rm SNR}\, \approxlt \, 12$.

\section{Notes to individual EZOA galaxies}  \label{app}



{\bf J1856$-$03:} This source was detected as HIZOA J1855$-$03B
by \citet{staveley16}: There is a possible galaxy faintly visible in
the NIR at (RA,Dec)\,=\,(18:56:00.5, $-03$:12:21) at a distance of
$2\farcm2$ to the \HI\ position. It is large and diffuse and, if real,
matches the profile well.

\noindent {\bf J1929$+$08: } Also detected as HIZOA J1929+08
by \citet{donley05};  their profile shows a clearly confused profile while
our (weaker) profile is rather noisy. The images show two galaxies of
similar size and appearance, both listed in the Deep Sky Hunters catalogue
by \citet{kronberger06}. DSH J1929.3+0802 is also an IRAS source.

\noindent {\bf J1919$+$14: } The \HI\ detection is affected by a strong
baseline variation which affects the fitted parameters. It was also
detected by \citet{donley05} at a distance of $5\farcm2$ to our \HI\
position. No obvious candidate could be found around either position but
the extinction is very high at $A_K=2\fm6$.

\noindent {\bf J1921$+$14: } This is an area of high extinction ($A_K = 1\fm7$,
variable across the search region) and star crowding. The UKIDSS images show
a diffuse detection consistent with the \HI\ parameters.

\noindent {\bf J2000$+$18: } The flux determination may be affected by an
RFI in the range $3900-4000$\,\kms\ which was not completely removed during
the reduction process.

\noindent {\bf J2001$+$26: } There are three galaxies in the field, one of
which (2MASX J20010969+2655338) was observed by \citet{kk18} with the \nrt\
but not detected. The other two galaxies are not listed in the literature:
the galaxy at (RA,Dec)\,=\,(20:01:26.4,+26:59:40) is diffuse and matches
the \HI\ parameters well, while the galaxy at
(RA,Dec)\,=\,(20:01:26.7,+26:55:36) at a larger distance of $3\farcm6$, is
an early spiral and similar in appearance to 2MASX J20010969+2655338.


\noindent {\bf J2029$+$31: } Despite being a high SNR detection (9.0), the
distance between the fitted position and the counterpart is rather large
($d$\,=\,$3\farcm4$). However, the position fitted by \sof\ is much closer
($d$\,=\,$0\farcm5$), and the diffuse appearance of the galaxy agrees well with
the \HI\ parameters.

\noindent {\bf J2050$+$47: }
Though only faintly visible on UKIDSS images, the galaxy is confirmed on
WISE images. 

\noindent {\bf J2102$+$46: }
There are two galaxies in the area, neither of which is listed in the
literature. The galaxy at (RA,Dec)\,= \,(21:02:33.9,+46:05:52), which is
faintly visible with WISE, appears to be larger and matches the \HI\
parameters better than the galaxy at (RA,Dec)\,=\,(21:02:49.9,+46:09:05).

\noindent {\bf J2130$+$48: }
There are two large galaxies in the field: 2MASX J21305323+4813559 at
$d$\,=\,$1\farcm5$ and 2MASX J21310014+4814279 at $d$\,=\,$1\farcm6$. Both were
observed by \citet{kk18} with the \nrt\ which confirmed
that 2MASX J21310014+4814279 is the counterpart. 2MASX J21305323+4813559
has also an optical velocity of 3556\,\kms\ \citep{seeberger98}.
 
\noindent {\bf J2131$+$43: }
The images show six galaxies, four of which form a tight group (with
possible interaction) at a rather large distance from our \HI\ position,
$d$\,=\,$3\farcm1$, and listed in the 2MASX catalogue as a single detection,
2MASX J21312321+4336182. It is also an IRAS source (IRAS 21295+4323) with
an optical velocity of $v=5467\pm56$\,\kms\ \citep{strauss92b}. There are
two \HI\ observations of this group (\citealt{paturel03}, 
\citealt{masters14}), both of which show our profile ($5500 - 5750$) as
well as a low-velocity shoulder which can be seen in the EBHIS cube as a
possible, fainter detection near-by and which is likely due to galaxies in
this group. The cross-match to our detection is most likely an unlisted
galaxy at (RA,Dec)\,=\,(21:31:17.9,+43:39:03) and $d$\,=\,$0\farcm4$. There
is another, fainter galaxy visible at (RA,Dec)\,=\,(21:31:16.9,+43:38:01),
$d$\,=\,$1\farcm4$. which may also belong to the group.

\noindent {\bf J2143$+$46: }
This \HI\ detection is likely confused. There are two large galaxies in the
field: UGC 11802 at $d$\,=\,$0\farcm5$ and UGC 11806 at
$d$\,=\,$3\farcm1$. Both are IRAS sources and of similar size and
morphological type, but UGC 11806 is more inclined. Optical velocities are
$v=3100\pm18$\,\kms\ and $v=3273\pm18$\,\kms\ \citep{karachentsev80},
respectively. Both have also been observed by \citet{theureau98b} with
the \nrt\ and were detected at $v=3153$\,\kms\ and 3322\,\kms ,
respectively.

\noindent {\bf J2237$+$53: }
There are two galaxies in the field, both of which contribute to the \HI\
detection. Both are of similar size but 2MASX J22370662+5357049
($d$\,=\,$0\farcm4$) is nearly edge on while 2MASX J22370933+5358339
($d$\,=\,$1\farcm2$) is fairly inclined and detected with IRAS. The former
was observed by \citet{seeberger94} with the Effelsberg radio telescope
with a similar profile to ours. The latter was observed
by \citet{courtois09} and \citet{paturel03} with the Green Bank 300-foot telescope
(GBT) and the \nrt , respectively. The GBT profile in particular shows
clearly two double horns with a low shoulder at $v<5400$\,\kms\ 
too faint to be visible in the EBHIS data.

\noindent {\bf J0020$+$59: }
This \HI\ detection is the close-by dwarf IC\,10. The extinction at this
position is overestimated since at latitudes $|b|<5\deg$ IRAS sources were
not all removed from the DIRBE/IRAS extinction maps \citep{schlegel98} 
and this galaxy is infrared-bright. It is also resolved with respect to
the telescope beam. More details are given in Sec.~\ref{ic10}.


\noindent {\bf J0213$+$66: }
There are two galaxies of similar appearance in the field, one is slightly
rounder than the other. The \HI\ detection shows no indication of
confusion, and it is not possible to decide which of the galaxies is the
counterpart.

\noindent {\bf J0233$+$58: }
The \HI\ detection shows a narrow single-peak profile; there is a galaxy
pair in the field consisting of one larger, elongated galaxy and one
smaller, roundish companion. The extinction is $A_K=0\fm3$ and no deep NIR
images are available, hence it is not possible decide if one of the
galaxies is nearly face-on. The larger galaxy was detected by IRAS and is
thus more likely to be the counterpart.

\noindent {\bf J0243$+$59: }
This detection is Maffei 2 and heavily affected by Galactic \HI\
contamination. The position and \HI\ parameters are unreliable. The
position as determined by \sof , (RA,Dec)\,=\,(02:42:03.3,+59:37:25), is
much closer to the galaxy's position ($d$\,=\,$1\farcm5$).

\noindent {\bf J0252$+$62: }
The distance between the \HI\ position and the counterpart
($d$\,=\,$4\farcm0$) is larger than could be expected for an SNR of 9.3. An
inspection of the cube shows a possible contamination at lower Galactic
latitude which may be a possible detection of a very LSB galaxy at
(RA,Dec)\,=\,(02:51:53.2,+62:27:43) and $d$\,=\,$4\farcm4$.

\noindent {\bf J0253$+$55: }
Outside the nominal search radius is a prominent galaxy pair: 2MASX
J02531475+5528143 at $d$\,=\,$4\farcm3$ and 2MASX J02531969+5529140 at
$d$\,=\,$4\farcm6$. \citet{paturel03} have observed the former while
\citet{kk18} have observed the latter, both with the NRT and showing similar
profiles, but only measuring half the flux of our detection. The galaxy
pair can therefore be ruled out as the counterpart. Instead, at
$d$\,=\,$2\farcm1$ an unpublished LSB is visible which matches the \HI\
parameters well.

\noindent {\bf J0255$+$57: }
The DSS images show a very faint detection at
(RA,Dec)\,=\,(02:55:16.9,+56:59:24) which is confirmed on the WISE
images; there are no deep NIR images.

\noindent {\bf J0308$+$62: }
The counterpart is only faintly visible on DSS and WISE images; there are
no deep NIR images.

\noindent {\bf J0314$+$64: }
This \HI\ detection is located at the edge of the cube and position
and \HI\ parameters are uncertain.


\noindent {\bf J0332$+$58: }
The detection is a close-by dwarf galaxy ($v=1430$\,\kms , \lmhi $ = 8.7$).
The UKIDSS images show a large, diffuse emission at
(RA,Dec)\,=\,(03:32:48.4,+58:14:55) which, if confirmed, is the
counterpart.

\noindent {\bf J0345$+$49: }
Though the distance of $4\farcm5$ between the \HI\ position and the
counterpart is rather large for an SNR of 7.4, at the low extinction of
$A_K=0\fm3$ and the large \HI\ mass of \lmhi $ = 9.8$ we can expect to see
the counterpart, and 2MASX J03455024+4914144 matches the \HI\ parameters
well.


\noindent {\bf J0437$+$54: }
There are two bright galaxies within the search radius: 2MASX
J04373506+5414339 at $d$\,=\,$1\farcm5$ seems to be little inclined, while
2MASX J04374087+5415389 at $d$\,=\,$2\farcm7$ is highly inclined; optical
redshifts are $\vopt=5369$\,\kms\ and $\vopt=5653$\,\kms \citep{huchra12},
respectively, compared to our $\vhel=5341$\,\kms . The latter galaxy was
observed by \citet{paturel03} with the \nrt\ at $v=5356$\,\kms , but
judging by the optical redshifts it is likely they detected 2MASX
J04373506+5414339 instead which lies well in the NRT beam.

\noindent {\bf J0437$+$43: }
Two large galaxies are visible within the search radius. The galaxy at
$d$\,=\,$1\farcm4$ is UGC\,03098 with an optical redshift of
$\vopt=4287\pm150$\,\kms\ \citep{hill88}, while the companion, 2MASX
J04370607+4355349 at $d$\,=\,$0\farcm8$, is an IRAS detection and appears
disturbed. UGC\,03098 was detected by \citet{springob05} with the GBT and
by \citet{staveley87} with PKS. Both show a lopsided profile similar to
ours which may be confused. It is not possible to tell which galaxy
contributes how much to the profile.

\noindent {\bf J0438$+$44: }
UGC\,03108, at $d$\,=\,$4\farcm4$ from the \HI\ position, was observed by
\citet{springob05} with the GBT and detected at $v=3959$\kms .
Their profile is wider than ours and shows our profile as a lopsided
high-velocity horn. We do not see the wider profile in our data due to the
lower sensitivity of the EBHIS survey. 

\noindent {\bf J0440$+$49A and J0440$+$49B: }
There are two barely separated \HI\ detections visible in the \HI\ cube:
the profile of J0440$+$49B covers $3750 - 3980$\,\kms\ and has a reliable
$W_{50}$, while J0440$+$49A starts at $v\simeq3600$\kms\ and goes also up
to $v\simeq3980$\kms . Its \wtw\ is close to the full line width of this
galaxy (bar the fact that the peak is dominated by the J0440$+$49B
profile). The images show a galaxy pair with one strongly inclined (broad
line width) and the other nearly face-on (narrow line width).

\noindent {\bf J0446$+$44: }
Though there are many galaxies visible on the UKIDSS images, all seem too
small for a velocity of $6434$\,\kms\ and \HI\ mass of \lmhi $ = 9.8$. They
are all likely part of a cluster at higher redshift.

\noindent {\bf J0455$+$34: }
The cross-match looks asymmetric and disturbed which makes it difficult to
estimate whether the galaxy could lie at a redshift of $v=3256$\,\kms . 

\noindent {\bf J0506$+$31: }
The cross-match is faint and diffuse, located near the edge of the UKIDSS
image. WISE images confirm this to be a galaxy. Another WISE galaxy, WISEP
J050601.69+314037.8 at $d$\,=\,$1\farcm3$. is likely further away 
\citep{rebull11}. 

\noindent {\bf J0520$+$43: }
There are two large galaxies in the field, possibly a pair. Our
cross-match, 2MASX J05200866+4314313, is an early type galaxy while the
other galaxy (2MASX J05201415+4318214 at $d$\,=\,$3\farcm2$) is a medium
type spiral. The former galaxy was detected with a similar profile
by \citet{kk18}, while the latter was observed by
\citet{takata94} with different \HI\ parameters.

\noindent {\bf J0546$+$31: }
One large galaxy and at least four smaller ones are visible on the
images, possibly a galaxy group. The largest is the most likely counterpart. 

\noindent {\bf J0554$+$18: }
The images show two large early-type galaxies and several smaller ones,
possibly part of a galaxy group. The most likely counterpart is 2MASX
J05540715+1759352, a face-on early-type spiral which shows strong 
star formation on WISE images. 



\noindent {\bf J0607$+$16: }
Well inside the search area lies a galaxy pair: 2MASX J06074379+1608036 at
$d$\,=\,$1\farcm4$ (inclined, barred spiral) and 2MASX J06074754+1604526 at
$d$\,=\,$2\farcm0$ (edge-on spiral). The former was detected
by \citet{kk18} with the NRT as clearly confused (and more lopsided than
our profile). 2MASX J06074754+1604526 was observed with Parkes by Said
(priv.\ comm.) which shows steeper edges and a pronounced peak at
$v\simeq5600$\,\kms\ as opposed to the NRT profile, indicating that both
galaxies contribute to our profile.

\noindent {\bf J0620$+$20: }
Two diffuse galaxies can be seen within the search radius; both were
detected in the Are\c{c}ibo Dual Beam Survey (ADBS, \citealt{rosenberg00})
at $v=1320$\,\kms (same as our detection) and at $v=2272$\,\kms\ (too faint to
be detected by us), respectively.

\noindent {\bf J0649$+$09: }
This galaxy was also detected by \citet{donley05} with PKS.  The most
likely cross-match is 2MASX J06493148+0939437 at $d$\,=\,$3\farcm3$. A very
diffuse late-type galaxy that is closer to the \HI\ position
($d$\,=\,$0\farcm3$) is less likely but cannot be excluded as counterpart;
it is possible that it contributes to the profile (not obvious, though) and
thus explains the larger distance to 2MASX J06493148+0939437.

\noindent {\bf J0636$+$00: }
There is a close galaxy pair in the field: 2MASX J06362668+0055433 is an
inclined spiral (with a possible small companion to the north) while 2MASX
J06362361+0055513 appears more face-on. The profile does not show an
indication of confusion (though it cannot be excluded), and the inclined
spiral, which is also an IRAS source, seems to be the more likely
counterpart.

\noindent {\bf J0700$-$04:} 
Begum \etal\ (2005) observed this galaxy with the VLA and found that
the \HI\ emission comes from two dwarf galaxies. Both are faintly visible
on the deep NIR images. See Sec.~\ref{ss3} for more details.

\bsp	
\label{lastpage}
\end{document}